\documentclass[a4paper,12pt]{article}
\pdfoutput=1
\usepackage[utf8]{inputenc}
\usepackage{amsmath}
\usepackage{amssymb}
\usepackage{mathrsfs}
\usepackage{authblk}
\usepackage[unicode,breaklinks=true]{hyperref}
\allowdisplaybreaks[1]

\title{Calculation of multipole moments of axistationary electrovacuum spacetimes}

\author[1]{Gyula Fodor}
\author[2]{Etevaldo dos Santos Costa Filho}
\author[2,3,4]{Betti Hartmann}

\affil[1]{Wigner Research Centre for Physics, 1525 Budapest 114, P.O.~Box 49, Hungary}
\affil[2]{Instituto de Física de São Carlos, Universidade de São Paulo, São Carlos, São Paulo, 13560-970, Brazil}
\affil[3]{Institut für Physik, Carl-von-Ossietzky Universität Oldenburg, 26111 Oldenburg, Germany}
\affil[4]{Department of Physics and Earth Sciences, Jacobs University Bremen, 28759 Bremen, Germany}

\begin{document}

\maketitle

\begin{abstract}
The multipole moments of stationary axially symmetric vacuum or electrovacuum spacetimes can be expressed in terms of the power series expansion coefficients of the Ernst potential on the axis. In this paper we present a simpler, more efficient calculation of the multipole moments, applying methods introduced by Bäckdahl and Herberthson. For the non-vacuum electromagnetic case, our results for the octupole and higher moments differ from the results already published in the literature. The reason for this difference is that we correct an earlier unnoticed mistake in the power series solution of the Ernst equations. We also apply the presented method to directly calculate the multipole moments of a 5-parameter charged magnetized generalization of the Kerr and Tomimatsu-Sato exact solutions.
\end{abstract}

\section{Introduction}

Multipole moments are important tools for the characterization of stationary spacetimes representing the exterior region of neutron stars or black holes. The purpose of this paper is the presentation of an improved, more efficient method for the calculation of the multipole moments of stationary axially symmetric spacetimes, when the moments are expressed in terms of the power series expansion coefficients of the Ernst potential on the axis of rotation. For the case when electromagnetic fields are also allowed, we give the correct expressions for the octupole and higher moments, correcting a mistake in the literature. As an application of the method, we calculate the multipole moments of a charged magnetized generalization of the Kerr and Tomimatsu-Sato solutions.

Multipole moment tensors for asymptotically flat static vacuum spacetimes were introduced in 1970 by Geroch, in a coordinate system independent way \cite{Geroch70}. The generalization of the definition for stationary spacetimes have been given by Hansen \cite{Hansen74}. In the stationary case there are two sets of multipole tensors, the mass moments and the angular momentum (or mass-current) moments, which can be unified into a set of complex valued quantities. Alternative, but equivalent definitions in terms of specific coordinate systems have been proposed by Thorne \cite{Thorne80,Gursel83}, and also by Simon and Beig \cite{Simon83}. A good review of early results on the topic can be found in \cite{Quevedo90}.

The concrete physical applications of multipole moments in general relativity have been pioneered by the work of Fintan D. Ryan \cite{Ryan95}. The gravitational radiation emitted by a compact object orbiting around a much larger central object can be used to determine the multipole moments of the central body. In case of extreme-mass-ratio inspirals, one is expected to be able to determine the first few gravitational multipole moments by the proposed space-based gravitational wave detector LISA \cite{Ryan97,LiLovelace08,Barack07,Babak17}.
Since the multipole moments of the Kerr black hole are uniquely determined by the mass and the angular momentum, this should provide a practical way of testing the no-hair theorem \cite{Chrusciel2012,Cardoso16,Cardoso19}.
Multipole moments can also be defined in alternative gravitational theories \cite{Sopuerta09,Pappas15a,Kleihaus14,Pappas15b}.
The multipole moments of so-called bumpy black holes and the gravitational radiation of test bodies orbiting around them have been studied in \cite{Collins04,Glampedakis06,Vigeland10}.
We expect that astrophysical observations will provide a way in the near future to test general relativity in the strong field regime \cite{Gair13,Berti15,Yagi16}.
In addition to gravitational waves, the observation of large compact objects and their multipole moments may be achieved by other methods, such as measuring the motion of stars or pulsars around them, study of accretion disks, or the observation black hole shadows by the event horizon telescope \cite{Will08,Broderick14,Suvorov16,Psaltis16}.

Further highly relativistic physical systems where multipole moments are important are neutron stars. In that case the observation of the spacetime structure outside the star is expected to give information about the equation of state of the matter in the interior \cite{Paschalidis17,Maselli20}. The innermost stable circular orbit marks the inner edge of accretion disks. Its properties, in terms of the multipole moments, has been calculated in \cite{Shibata98,SanabriaGomez10,Berti04}. It is possible to find certain universal relations between the multipole moments of a neutron star, establishing a no-hair property for these objects. The three-hair relations determine the higher multipole moments by the mass, angular momentum and quadrupole moment, in an approximately equation of state independent way \cite{Pappas14,Yagi14,Yagi17}. Exact solutions for the vacuum exterior region have played important role in the establishment of these universal relations \cite{Manko95,Manko00,Manko00b,Teichmuller11,Pachon12,Pappas13,Manko16}.

The above physical applications have been made possible by strict mathematical results establishing the theory of multipole moments for the nonlinear Einstein equations in the stationary case. Multipole moments are defined for any asymptotically flat stationary spacetime, and if two spacetimes have the same multipole moments, then they agree at least in a neighborhood of conformal infinity \cite{Xanthopoulos79,Beig80,Kundu81,Beig81}.

Most astrophysically relevant nonradiating spacetimes are expected to be axially symmetric. In the axisymmetric case the multipole moment tensor of order $n$ can be represented by a single scalar moment, called $P_n$ by Hansen \cite{Hansen74}. These scalar moments can be expressed in terms of the power series expansion coefficients of the Ernst potential on the symmetry axis \cite{Hoenselaers86}. An algorithm for this calculation has been published in 1989 by Fodor, Hoenselaers and Perjés \cite{Fodor89}. Since the calculation of the Ernst potential on the axis is relatively easy, the results presented in \cite{Fodor89} became a standard tool for obtaining the multipole moments. The method has been applied not only for exact solutions, but also for gravitational radiation \cite{Ryan95}, innermost circular orbits \cite{Shibata98}, and neutron stars \cite{Yagi14}. Multipole moments of a rigidly rotating disk of dust has been calculated in \cite{Kleinwachter95}. If the axistationary spacetime is reflection symmetric with respect to the equatorial plane, then for even $n$ the moments $P_n$ are real, while for odd $n$ they are purely imaginary \cite{Kordas95,Meinel95}.

Further important developments on the mathematical theory of multipole moments can be found in a series of papers published by B\"akdahl and Herberthson. A major aim in their considerations is the proof of a long-standing conjecture by Geroch \cite{Geroch70}. This conjecture claims that if one chooses any set of multipole moments that satisfy some appropriate convergence conditions, then there always exists a spacetime having precisely those moments. This has been proven first for the static axially symmetric case \cite{Herberthson04,Backdahl05a}, and then for stationary axially symmetric spacetimes \cite{Backdahl05,Backdahl07}. Stationary case without the assumption of axisymmetry have been considered in \cite{Backdahl06}, with a proof of the necessary part of the conjecture. Finally a proof for the general static case has been given in \cite{Herberthson09}.

B\"akdahl and Herberthson also introduce some very useful tools that make the calculation of multipole moments considerably simpler. They define a complex null vector field, which makes the operation of taking the symmetric and trace-free part of tensors simple and trivial. They also introduce the concept of leading order part of functions, which allows the use of functions depending on only one variable instead of two. As far as we know, these tools have not been used yet for the calculation of multipole moments of exact solutions, apart from the Kerr case. In this paper we will re-calculate the results given in \cite{Fodor89}, where the moments are given in terms of the Ernst potential on the axis, using the methods of B\"akdahl and Herberthson. We will also look at the electromagnetic generalization of this procedure.

Multipole moments for stationary Einstein-Maxwell fields have been defined by Simon \cite{Simon84}. In this case there are two sets of complex multipole moment tensors, and there are also two complex Ernst potentials. Conditions on equatorial symmetry or antisymmetry for stationary axisymmetric electrovacuum spacetimes have been discussed in \cite{Pachon06,Ernst06}.

For the axially symmetric electrovacuum case, the procedure for calculating the multipoles in terms of the axis coefficients of the Ernst potentials have been published first in \cite{Hoenselaers90}. Unfortunately, there have been two mistakes in that paper, which also affected the end results for the multipole moments. The first mistake has been found and corrected in \cite{Sotiriou04}. However, a second mistake remained unnoticed, which has been pointed out only in a conference proceedings article \cite{Perjes03}. In the present paper we again calculate the moments, now using the simpler method of B\"akdahl and Herberthson. We give the correct expressions for the scalar gravitational multipole moments $P_n$ and electromagnetic moments $Q_n$ up to order $n=6$. Higher order moments can be easily obtained by the Mathematica or Maple file provided as supplementary material. 

Denoting the expansion coefficients of the gravitational and electromagnetic Ernst potentials by $m_n$ and $q_n$, respectively, for the first three moments we obtain the expected result:
\begin{equation}
 P_n=m_n \quad , \qquad Q_n=q_n \qquad \mathrm{for} \qquad n=0 , 1 , 2 \ .
\end{equation}
For the vacuum case it is known that $P_3=m_3$, but if there are electromagnetic fields, then we obtain that generally $P_3\not=m_3$ and $Q_3\not=q_3$. This is a clear difference from earlier results published in \cite{Hoenselaers90,Sotiriou04}, where the differences started only from $n=4$. Actually, the mistake has not been in the calculation of the moments, that is the same procedure for the vacuum and electrovacuum case, but in the power series solution of the Ernst equations. We find it important to give the correct expressions for the power series solution and for the moments in the electromagnetic case, since these results have been used in several subsequent papers. In the not too far future astrophysical observations may become precise enough to make the octupole moment a measurable quantity.

In the last section of the paper we apply the earlier discussed methods for the calculation of the multipole moments of a five-parameter exact solution presented in \cite{Manko00b}, which is a charged magnetized generalization of both the Kerr and the $\delta=2$ Tomimatsu-Sato solutions. The solution is general enough to describe both sub-extreme and hyper-extreme configurations, and the expressions that we obtain for the multipole moments are valid for both cases. We print the moments up to order $n=5$, but we provide an algebraic manipulation software code as supplementary material, to allow higher order calculations.

The structure of the paper is the following. In Section \ref{secstacgen} we give a short review on how multipole moments can be defined for general stationary spacetimes, when electromagnetic fields can be present. Here we fix notations, sign-conventions, and present the field equations to be solved. In Section \ref{secaxisimm} we specialize to axially symmetric solutions, and present the theory needed for the definition of the scalar multipole moments using Weyl coordinates. Here we also discuss those tools and methods introduced by B\"akdahl and Herberthson which may be useful for the calculation of the moments of exact solutions. In Section \ref{secexpmulti} we calculate the gravitational and electromagnetic multipole moments in terms of the expansion coefficients of the Ernst potentials on the axis. By listing the results, we correct a mistake that remained unnoticed in the literature for quite many years. Finally, in Section \ref{secmagtssol} we directly apply the B\"akdahl-Herberthson method to calculate the multipole moments of an exact solution, which is general enough to approximate well the exterior region of rotating neutron stars \cite{Manko00,Manko00b,Berti05}.

\section{Stationary electrovacuum spacetimes} \label{secstacgen}

\subsection{Ernst equations}

The complex Ernst potential $\mathcal{E}$ was introduced initially for axisymmetric stationary spacetimes by Ernst in 1968. It was first presented for the vacuum case \cite{Ernst68a}, and then also for electrovacuum by adding a second complex potential $\Phi$ describing the electromagnetic field \cite{Ernst68b}. Subsequently, it was shown that these potentials can also be defined for general stationary electrovacuum spacetimes \cite{Harrison68,Ernst71,Israel72,Kinnersley73}.

We solve the Einstein equations $G_{\mu\nu}=8\pi T_{\mu\nu}$ with the electromagnetic stress tensor
\begin{equation}
 T_{\mu\nu}=\frac{1}{4\pi}\left(F_{\mu\rho}F_{\nu}^{\ \rho}
 -\frac{1}{4}g_{\mu\nu}F_{\rho\sigma}F^{\rho\sigma}\right) \ . \label{eqtmunuff}
\end{equation}
Since $T_{\mu\nu}$ is traceless, the Ricci scalar is necessarily zero. Introducing the vector potential $A_\mu$ by
\begin{equation}
 F_{\mu\nu}=\partial_\mu A_\nu-\partial_\nu A_\mu \ , \label{eqfmunuaa}
\end{equation}
the remaining Maxwell's equation is $\nabla_\mu F^{\mu\nu}=0$. The choice in the order of indices in \eqref{eqfmunuaa} is not important, since the Einstein and Maxwell equations are invariant under $F_{\mu\nu} \to -F_{\mu\nu}$. The Ricci tensor definition we use is
\begin{equation}
 R_{\mu\nu}=\partial_\rho\Gamma^\rho_{\ \mu\nu}-\partial_\mu\Gamma^\rho_{\ \rho\nu}
 +\Gamma^\rho_{\ \mu\nu}\Gamma^\sigma_{\ \sigma\rho}
 -\Gamma^\rho_{\ \sigma\mu}\Gamma^\sigma_{\ \rho\nu} \ . \label{eqrrmunugam}
\end{equation}
We assume that the metric signature is $(-,+,+,+)$. Replacing $g_{\mu\nu}$ by $-g_{\mu\nu}$ leaves the Christoffel symbols and the Ricci tensor invariant, however, $T_{\mu\nu}$ changes sign. Hence in case of signature choice $(+,-,-,-)$ a factor $-1$ in \eqref{eqtmunuff}, or in the Einstein equation $G_{\mu\nu}=8\pi T_{\mu\nu}$, or possibly in \eqref{eqrrmunugam} would be necessary. All the equations and definitions in the following are identically same for the two signature choices, except for those determining the four-dimensional spacetime metric.

Since we consider stationary spacetimes, we use tensorial quantities defined on the three-manifold $\mathcal{M}$ of the trajectories of the timelike Killing vector $\xi^\mu$ \cite{Geroch71}. Denoting the spacetime metric by $g_{\mu\nu}$, and the norm of the Killing vector by $f=-\xi^\mu \xi_\mu>0$, on the trajectories of $\xi^\mu$ we define the rescaled induced metric as
\begin{equation}
 h_{\mu\nu}=f g_{\mu\nu}+\xi_\mu\xi_\nu \ .
\end{equation}
For the three-dimensional tensors we use Latin indices, which are raised and lowered by the metric $h_{ab}$. The derivative operator belonging to $h_{ab}$ is denoted by $\nabla_a$. Using a coordinate system adapted to the timelike Killing vector, the spacetime metric can be written as
\begin{equation}
 \mathrm{d}s^2=-f\left(\mathrm{d}t+\omega_a\mathrm{d}x^a\right)^2
 +\frac{1}{f}h_{ab}\mathrm{d}x^a\mathrm{d}x^b \ , \label{eqdsstacgen}
\end{equation}
where $f$, $\omega_a$ and $h_{ab}$ are independent of $t$. For stationary electromagnetic fields the complex electromagnetic Ernst potential $\Phi$ can be defined in terms of the four-dimensional vector potential $A_\mu=(A_t,A_a)$ as \cite{Israel72,Simon84}
\begin{equation}
 \Phi=A_t+i A' \ , \label{eqphiaa}
\end{equation}
where the real scalar $A'$ is determined by
\begin{equation}
 \nabla_a A'=f \epsilon_{abc}\left(\nabla^b A^c
 +\omega^b\nabla^c A_t\right) \ , \label{eqnablaap}
\end{equation}
$\epsilon_{abc}=\sqrt{h}\,\varepsilon_{abc}$ is the three-dimensional Levi-Civita tensor, and the spatial indices of $A_\mu$ has been raised by $h^{ab}$. For the electromagnetic case the complex Ernst potential is defined as \cite{Israel72,Simon84}
\begin{equation}
 \mathcal{E}=f+i\chi-\Phi\overline\Phi \ , \label{eqeefipsi}
\end{equation}
where
\begin{equation}
 \nabla_a\chi=f^2\epsilon_{abc}\nabla^b\omega^c
 +i\left(\overline\Phi\nabla_a\Phi-\Phi\nabla_a\overline\Phi\right) \ , \label{eqnablapsi}
\end{equation}
and overline denotes complex conjugation. Obviously,
\begin{equation}
 f=\frac{1}{2}\left(\mathcal{E}+\overline{\mathcal{E}}\right)
 +\Phi\overline\Phi \ .
\end{equation}

The Einstein and Maxwell equations are equivalent to the following three equations \cite{Israel72,Kinnersley73,Stephani03}:
\begin{align}
 f\Delta\mathcal{E}=&\nabla^a\mathcal{E}\left(\nabla_a\mathcal{E}
 +2\overline\Phi \nabla_a\Phi\right) \ , \label{eqfdeleps}\\
 f\Delta\Phi=&\nabla^a\Phi\left(\nabla_a\mathcal{E}
 +2\overline\Phi \nabla_a\Phi\right) \ , \label{eqfdelphi}\\
 f^2 R_{ab}=&\frac{1}{2}\nabla_{(a}\mathcal{E}\nabla_{b)}\overline{\mathcal{E}}
 +\Phi \nabla_{(a}\mathcal{E}\nabla_{b)}\overline{\Phi}
 +\overline\Phi \nabla_{(a}\overline{\mathcal{E}}\nabla_{b)}\Phi \notag\\
 &-\left(\mathcal{E}+\overline{\mathcal{E}}\right)
 \nabla_{(a}\Phi \nabla_{b)}\overline{\Phi} \ , \label{eqfrrab}
\end{align}
where $\Delta=\nabla^a\nabla_a$ is the Laplacian, and $R_{ab}$ is the Ricci tensor belonging to three-dimensional metric $h_{ab}$. The unknowns in these equations are the two complex potentials $\mathcal{E}$ and $\Phi$, together with the three-metric components $h_{ab}$.

We introduce the complex potentials
\begin{equation}
 \xi=\frac{1-\mathcal{E}}{1+\mathcal{E}} \quad , \qquad
 q=\frac{2\Phi}{1+\mathcal{E}} \ , \label{eqxiqdef}
\end{equation}
with inverse relations
\begin{equation}
 \mathcal{E}=\frac{1-\xi}{1+\xi} \quad , \qquad
 \Phi=\frac{q}{1+\xi} \ .
\end{equation}
We note that the above $\xi$ is the usual definition in the papers on multipole moments \cite{Fodor89,Hoenselaers90,Sotiriou04}, while it is the inverse of the $\xi$ originally defined by Ernst in \cite{Ernst68a,Ernst68b}.
In terms of these potentials the Einstein-Maxwell equations \eqref{eqfdeleps}-\eqref{eqfrrab} can be written into the form
\begin{align}
 \Theta\Delta\xi=&2\left(\overline\xi\nabla^a\xi-\overline q\nabla^a q\right)
 \nabla_a\xi \ , \label{eqthdelxi} \\
 \Theta\Delta q=&2\left(\overline\xi\nabla^a\xi-\overline q\nabla^a q\right)
 \nabla_a q \ , \label{eqthdelq}\\
 \Theta^2 R_{ab}=&2\,\mathrm{Re}\left(
 \nabla_a\xi\nabla_b\overline\xi-\nabla_a q\nabla_b\overline q
 +s_a \overline s_b
 \right) \label{eqthrrab}\ ,
\end{align}
where
\begin{align}
 \Theta&=\xi\overline\xi-q\overline q-1 \ , \\
 s_a&=\xi\nabla_a q-q\nabla_a\xi \ .
\end{align}

\subsection{Asymptotic flatness} 

According to the definition of Penrose and Geroch \cite{Penrose65,Geroch70}, a 3-dimensional manifold $\mathcal{M}$ with positive definite metric $h_{ab}$ is \emph{asymptotically flat} if the following conditions hold:
\begin{enumerate}
 \item There exists a manifold $\tilde{\mathcal{M}}$ with metric $\tilde h_{\tilde a\tilde b}$ and a diffeomorphism ${\psi:\mathcal{M}\to\tilde{\mathcal{M}}\setminus\Lambda}$, where $\Lambda$ is a single point in $\tilde{\mathcal{M}}$, such that $\psi$ is a conformal isometry with conformal factor $\Omega$, i.e.~$\tilde h_{\tilde a\tilde b}=\Omega^2(\psi^{*}h)_{\tilde a\tilde b}$ \ .
 \item The function $\Omega$ can be extended as a $C^2$ scalar to the point $\Lambda$ corresponding to spatial infinity, such that
\begin{equation}
 \Omega\big|_\Lambda=0 \ , \quad \tilde\nabla_{\tilde a}\Omega\big|_\Lambda=0 \ , \quad \tilde\nabla_{\tilde a}\tilde\nabla_{\tilde b}\Omega\big|_\Lambda=2\tilde h_{\tilde a\tilde b}\big|_\Lambda \ ,
 \label{eqomlambda}
\end{equation}
where $\tilde\nabla_{\tilde a}$ is the derivative operator on $\tilde{\mathcal{M}}$ belonging to $\tilde h_{\tilde a\tilde b}$.
\end{enumerate}
We use tilde on the coordinate indices to indicate that these are tensors on $\tilde{\mathcal{M}}$, and the coordinate system used on that manifold is generally different from the mapped version of the original coordinates on $\mathcal{M}$.

\subsection{Multipole moments} \label{secgenmultipole}

Let us choose a real or complex scalar field $\phi$ on $\mathcal{M}$, and assume that $\tilde\phi=\Omega^{-1/2}\phi$ can be smoothly extended to the point $\Lambda$. We define a set of tensor fields recursively \cite{Geroch70,Hansen74,Simon84},
\begin{align}
 \mathcal{P}^{(0)}&=\tilde\phi \ , \label{eqpnp1ind0}\\
 \mathcal{P}^{(1)}_{\tilde a}&=
 \tilde\nabla_{\tilde a}\mathcal{P}^{(0)} \ , \\
 \mathcal{P}^{(n+1)}_{\tilde a_1\ldots\tilde a_{n+1}}&=
 \mathscr{C}\left[
 \tilde\nabla_{\tilde a_{n+1}}\mathcal{P}^{(n)}_{\tilde a_1\ldots\tilde a_{n}}
 -\frac{1}{2}n(2n-1)\tilde R_{\tilde a_1\tilde a_2}
 \mathcal{P}^{(n-1)}_{\tilde a_3\ldots\tilde a_{n+1}}
 \right] \ , \label{eqpnp1ind}
\end{align}
where $\tilde R_{\tilde a\tilde b}$ is the Ricci tensor belonging to $\tilde h_{\tilde a\tilde b}$, and $\mathscr{C}$ denotes the operation of taking the symmetric trace-free part. For details on how to perform the operation $\mathscr{C}$ see e.g.~\cite{Trautman65}, \cite{Thorne80}, or the Appendix of \cite{Fodor89}. The multipole moment tensors are defined as the values of these tensor fields at infinity,
\begin{equation}
 M^{(n)}_{\tilde a_1\ldots\tilde a_{n}}=
 \mathcal{P}^{(n)}_{\tilde a_1\ldots\tilde a_{n}}\bigr{|}_\Lambda \ . \label{eqmmnppn}
\end{equation}
The choice $\phi=\xi$ gives the gravitational moment tensors $P^{(n)}_{\tilde a_1\ldots\tilde a_{n}}\equiv M^{(n)}_{\tilde a_1\ldots\tilde a_{n}}$, while the choice $\phi=q$ yields the electromagnetic moments $Q^{(n)}_{\tilde a_1\ldots\tilde a_{n}}\equiv M^{(n)}_{\tilde a_1\ldots\tilde a_{n}}$. The recursion relations \eqref{eqpnp1ind0}-\eqref{eqpnp1ind} do not mix the real and imaginary parts. Hence the moments calculated from a complex $\phi$ are equivalent to two sets of real moments. The real and imaginary parts of $P^{(n)}_{\tilde a_1\ldots\tilde a_{n}}$ are the mass and angular momentum moments respectively, while $Q^{(n)}_{\tilde a_1\ldots\tilde a_{n}}$ provides the electric and magnetic moments. The imaginary part of $P^{(n)}_{\tilde a_1\ldots\tilde a_{n}}$ are called current or mass-current moments in papers related to gravitational radiation \cite{Thorne80,Ryan95,Sotiriou05}, and this became the standard in recent literature.

The angular momentum monopole would correspond to a NUT charge, which must be zero in the asymptotically flat case \cite{Hansen74}. Hence the gravitational monopole moment has to be real, and it agrees with the mass of the system, $P^{(0)}\equiv M$. Similarly, the electromagnetic monopole moment has to be real, expressing the absence of magnetic monopoles \cite{Simon84}. The electric charge is given by $Q^{(0)}\equiv Q$.

According to \eqref{eqomlambda}, if we change the conformal factor as $\tilde\Omega=\tilde\omega\Omega$, the space will be asymptotically flat in terms of $\tilde\Omega$ if $\tilde\omega|_\Lambda=1$. In this case, the multipole moments transform according to the formula obtained by Beig \cite{Beig81a},
\begin{align}
 \tilde M^{(n)}_{\tilde a_1\ldots\tilde a_{n}}=&
 M^{(n)}_{\tilde a_1\ldots\tilde a_{n}} \label{eqbarmmna} \\
 &+\mathscr{C}\sum_{k=0}^{n-1}\binom{n}{k}\frac{(2n-1)!!}{(2k-1)!!}
 (-2)^{k-n}M^{(k)}_{\tilde a_1\ldots\tilde a_{k}}
 \left(\partial_{\tilde a_{k+1}}\tilde\omega\right)_{\Lambda}\ldots
 \left(\partial_{\tilde a_{n}}\tilde\omega\right)_{\Lambda}
 \ . \notag 
\end{align}
The change of the moments depend only on $\left(\partial_{\tilde a}\tilde\omega\right)_{\Lambda}$, which is a vector of only $3$ real components at a single point. The conformal rescaling generally corresponds to a spatial shift in the center of mass. The conformal transformation is usually used to make the real part of the gravitational dipole moment $P^{(1)}_{\tilde a}$ zero, ensuring a center of mass coordinate system. Setting $\tilde\omega|_\Lambda=1$ and $\left(\partial_{\tilde a}\tilde\omega\right)_{\Lambda}=0$, the higher derivatives of the function $\tilde\omega$ can be chosen arbitrarily, keeping the multipole moments invariant.

\section{Axisymmetric electrovacuum} \label{secaxisimm}

\subsection{Spacetime metric}

In the axially symmetric case, specializing \eqref{eqdsstacgen}, we write the metric into the Weyl-Lewis-Papapetrou form,
\begin{equation}
 \mathrm{d}s^2=-f\left(\mathrm{d}t-\omega\mathrm{d}\varphi\right)^2
 +\frac{1}{f}\left[e^{2\gamma}\left(\mathrm{d}\rho^2+\mathrm{d}z^2\right)
 +\rho^2\mathrm{d}\varphi^2\right] \ , \label{eqdsweyl}
\end{equation}
where $f$, $\omega$ and $\gamma$ are functions of the coordinates $\rho$ and $z$. In terms of the spatial coordinates $x^a=(\rho,z,\varphi)$ the metric on $\mathcal{M}$ is
\begin{equation}
 h_{ab}=\left(
 \begin{array}{ccc}
  e^{2\gamma} & 0 & 0 \\
  0 & e^{2\gamma} & 0 \\
  0 & 0 & \rho^2
 \end{array}
 \right) \ . \label{eqhabaxisym}
\end{equation}
The only nonvanishing component of $\omega_a$ in \eqref{eqdsstacgen} is now $\omega_\varphi\equiv -\omega$. At the rotation axis $\rho=0$ necessarily $\gamma=0$, because of the absence of conical singularity.

In axistationary spacetimes the vector potential has only two components, $A_t$ and $A_\varphi$, which depend on the coordinates $\rho$ and $z$. The potential $\Phi$ is defined by \eqref{eqphiaa}, and equation \eqref{eqnablaap} for $A'$ can be written as \cite{Ernst68b}
\begin{equation}
 \partial_\rho A'=\frac{f}{\rho}\left(\partial_z A_\varphi+\omega\partial_z A_t\right)
 \quad , \qquad
 \partial_z A'=-\frac{f}{\rho}\left(\partial_\rho A_\varphi+\omega\partial_\rho A_t\right) \ .
\end{equation}
The Ernst potential $\mathcal{E}$ is defined by \eqref{eqeefipsi}, where now \eqref{eqnablapsi} takes the form
\begin{align}
 \partial_\rho\chi&=-\frac{1}{\rho}f^2\partial_z\omega
 +i\left(\overline\Phi\partial_\rho\Phi-\Phi\partial_\rho\overline\Phi\right) \ , \\
 \partial_z\chi&=\frac{1}{\rho}f^2\partial_\rho\omega
 +i\left(\overline\Phi\partial_z\Phi-\Phi\partial_z\overline\Phi\right) \ .
\end{align}
These expressions determine the relation between the complex Ernst potentials and the spacetime metric components.

Choosing the spatial metric $h_{ab}$ in the form \eqref{eqhabaxisym}, for arbitrary axially symmetric functions $f$ the Laplacian is
\begin{equation}
 \Delta f=\frac{1}{\sqrt{h}}\partial_a\left(\sqrt{h}h^{ab}\partial_b f\right)
 =e^{-2\gamma}\left(\partial^2_\rho f+\frac{1}{\rho}\partial_\rho f
 +\partial^2_z f\right) \ . \label{eqlaplf1}
\end{equation}
For axially symmetric functions $f$ and $g$ the product of the derivatives is
\begin{equation}
 \nabla^a f\nabla_a g=h^{ab}\partial_a f\partial_b g=
 e^{-2\gamma}\left(\partial_\rho f\partial_\rho g
 +\partial_z f\partial_z g\right) \ . \label{eqnabgnabg1}
\end{equation}
This shows that the factor $e^{-2\gamma}$ drops out from the Ernst equations \eqref{eqthdelxi} and \eqref{eqthdelq}, giving a coupled system of equations for determining $\xi$ and $q$. These two relations can also be considered as equations on flat 3-dimensional space with metric $h_{ab}^{(0)}=\mathrm{diag}(1,1,\rho^2)$. The right-hand side of \eqref{eqthrrab} is metric independent, the derivatives can be substituted by partial derivatives. On the left-hand side, for our choice of coordinates we have
\begin{equation}
 R_{ab}=\left(
 \begin{array}{ccc}
  -\partial^2_\rho\gamma+\frac{1}{\rho}\partial_\rho\gamma-\partial^2_z\gamma
  & \frac{1}{\rho}\partial_z\gamma & 0 \\
  \frac{1}{\rho}\partial_z\gamma
  & -\partial^2_\rho\gamma-\frac{1}{\rho}\partial_\rho\gamma-\partial^2_z\gamma & 0 \\
  0 & 0 & 0
 \end{array}
 \right) \ , \label{eqricciab1}
\end{equation}
from which
\begin{equation}
 \partial_\rho\gamma=\frac{1}{2}\rho(R_{\rho\rho}-R_{zz}) \quad \ ,  \qquad
 \partial_z\gamma=\rho R_{\rho z} \ .
\end{equation}
After $\xi$ and $q$ are known from \eqref{eqthdelxi}-\eqref{eqthdelq}, these can be used together with \eqref{eqthrrab} to obtain $\gamma$.

\subsection{Asymptotic coordinates}

In order to describe the region near infinity, on the manifold $\mathcal{M}$ we introduce the coordinates
\begin{equation}
 \tilde\rho=\frac{\rho}{r^2} \quad ,
 \qquad \tilde z=\frac{z}{r^2} \ , \label{eqrhotilztil}
\end{equation}
where $r^2=\rho^2+z^2$. In terms of the coordinates $\tilde x^{\tilde a}=(\tilde\rho,\tilde z,\varphi)$ the three-dimensional metric becomes
\begin{equation}
 h_{\tilde a\tilde b}=\frac{1}{\tilde r^4}\left(
 \begin{array}{ccc}
  e^{2\gamma} & 0 & 0 \\
  0 & e^{2\gamma} & 0 \\
  0 & 0 & \tilde\rho^2
 \end{array}
 \right) \ , \label{eqhabaxisymbar}
\end{equation}
where $\tilde r^2=\tilde \rho^2+\tilde z^2=r^{-2}$.
%Since we will only work in the coordinate system $\tilde x^a$, we drop the overbar from $\tilde\rho$, $\tilde z$ and $\tilde r$ from now.
The field equations \eqref{eqthdelxi}-\eqref{eqthrrab} are still valid in this coordinate system, where now
\begin{align}
 &\Delta f
 =\tilde r^4e^{-2\gamma}\left(\partial^2_{\tilde\rho}f
 +\frac{1}{\tilde\rho}\partial_{\tilde\rho}f
 +\partial^2_{\tilde z}f-\frac{2\tilde\rho}{\tilde r^2}\partial_{\tilde\rho}f
 -\frac{2\tilde z}{\tilde r^2}\partial_{\tilde z}f
 \right) \label{eqlaplf2} \ ,  \\
 &\nabla^{\tilde a} f\nabla_{\tilde a}g=
 \tilde r^4e^{-2\gamma}\left(\partial_{\tilde\rho}f\partial_{\tilde\rho}g
 +\partial_{\tilde z}f\partial_{\tilde z}g\right) \ , \label{eqnabtildaf}
\end{align}
instead of \eqref{eqlaplf1} and \eqref{eqnabgnabg1}. The nonvanishing components of the Ricci tensor are
\begin{align}
 R_{\tilde\rho\tilde\rho}&=
  -\partial^2_{\tilde\rho}\gamma+\frac{1}{\tilde\rho}\partial_{\tilde\rho}\gamma
  -\partial^2_{\tilde z}\gamma
  -\frac{2\tilde\rho}{\tilde r^2}\partial_{\tilde\rho}\gamma
  +\frac{2\tilde z}{\tilde r^2}\partial_{\tilde z}\gamma \ , \label{eqrrtt11}\\
 R_{\tilde z\tilde z}&=-\partial^2_{\tilde\rho}\gamma
 -\frac{1}{\tilde\rho}\partial_{\tilde\rho}\gamma-\partial^2_{\tilde z}\gamma
  +\frac{2\tilde\rho}{\tilde r^2}\partial_{\tilde\rho}\gamma
  -\frac{2\tilde z}{\tilde r^2}\partial_{\tilde z}\gamma \ , \\
 R_{\tilde\rho\tilde z}&= \frac{1}{\tilde\rho}\partial_{\tilde z}\gamma
 -\frac{2\tilde z}{\tilde r^2}\partial_{\tilde\rho}\gamma
 -\frac{2\tilde\rho}{\tilde r^2}\partial_{\tilde z}\gamma \ . \label{eqrrtt22}
\end{align}
Since $\partial_{\tilde\rho}\tilde r=\frac{\tilde\rho}{\tilde r}$ and $\partial_{\tilde z}\tilde r=\frac{\tilde z}{\tilde r}$ the Laplacian \eqref{eqlaplf2} can also be written as
\begin{equation}
 \Delta f
 =\tilde r^5 e^{-2\gamma}\left(\partial^2_{\tilde\rho}\,\frac{f}{\tilde r}
 +\frac{1}{\tilde\rho}\partial_{\tilde\rho}\,\frac{f}{\tilde r}
 +\partial^2_{\tilde z}\,\frac{f}{\tilde r}
 \right) \label{eqlaplf3} \ .
\end{equation}

\subsection{Conformal mapping}

We use a specific conformal factor $\Omega=\tilde r^2$ to define the metric $\tilde h_{\tilde a\tilde b}=\Omega^2h_{\tilde a\tilde b}$. Using the $\tilde x^{\tilde a}=(\tilde\rho,\tilde z,\varphi)$ coordinates this metric has the form
\begin{equation}
 \tilde h_{\tilde a\tilde b}=\left(
 \begin{array}{ccc}
  e^{2\gamma} & 0 & 0 \\
  0 & e^{2\gamma} & 0 \\
  0 & 0 & \tilde\rho^2
 \end{array}
 \right) \ , \label{eqhabaxisyminf}
\end{equation}
which obviously can be smoothly extended to the point $\tilde\rho=\tilde z=0$, so it is a metric on $\tilde{\mathcal{M}}$. This point, denoted by $\Lambda$, corresponds to spatial conformal infinity. Since on the axis $\gamma=0$, the choice $\Omega=\tilde r^2$ obviously satisfies the conditions of asymptotic flatness given in \eqref{eqomlambda}.

The metric in \eqref{eqhabaxisyminf} has the same structure as in \eqref{eqhabaxisym}. Hence the Laplacian $\tilde\Delta f$ and the product $\tilde\nabla^{\tilde a}f\tilde\nabla_{\tilde a}g$ belonging to this new metric have the same form as in \eqref{eqlaplf1} and \eqref{eqnabgnabg1} respectively, in terms of the tilded coordinates. We define the rescaled potentials as
\begin{equation}
 \tilde\xi=\Omega^{-1/2}\xi=\frac{\xi}{\tilde r} \quad , \qquad 
 \tilde q=\Omega^{-1/2}q=\frac{q}{\tilde r} \ . \label{eqhatxihatq}
\end{equation}
Using \eqref{eqnabtildaf} and \eqref{eqlaplf3}, the Ernst equations \eqref{eqthdelxi}-\eqref{eqthdelq} can be written as
\begin{align}
 \Theta\tilde\Delta\tilde\xi=&2\left[
 \overline{\tilde\xi}\,\tilde\nabla^{\tilde a}(\tilde r\tilde\xi)
 -\overline{\tilde q}\,\tilde\nabla^{\tilde a}(\tilde r\tilde q)\right]
 \tilde\nabla_{\tilde a}(\tilde r\tilde\xi) \ , \label{eqthdelxi2} \\
 \Theta\tilde\Delta\tilde q=&2\left[
 \overline{\tilde\xi}\,\tilde\nabla^{\tilde a}(\tilde r\tilde\xi)
 -\overline{\tilde q}\,\tilde\nabla^{\tilde a}(\tilde r\tilde q)\right]
 \tilde\nabla_{\tilde a}(\tilde r\tilde q) \ , \label{eqthdelq2}
\end{align}
where $\Theta=\tilde r^2\tilde\xi\overline{\tilde\xi}-\tilde r^2\tilde q\overline{\tilde q}-1$.

We introduce the operators $\tilde D_{\tilde a}$ defined in \cite{Hoenselaers90,Sotiriou04}, such that
\begin{align}
 \tilde D_{\tilde\rho}\tilde\xi&=\tilde z\partial_{\tilde\rho}\tilde\xi
 -\tilde\rho\partial_{\tilde z}\tilde\xi
 \  ,
 &\tilde D_{\tilde z}\tilde\xi&=\tilde\rho\partial_{\tilde\rho}\tilde\xi
 +\tilde z\partial_{\tilde z}\tilde\xi+\tilde\xi
 \  ,
 &\tilde D_{\varphi}\tilde\xi&=0 \ , \label{eqtdtrhtxi} \\
 \tilde D_{\tilde\rho}\tilde q&=\tilde z\partial_{\tilde\rho}\tilde q
 -\tilde\rho\partial_{\tilde z}\tilde q
 \  ,
 &\tilde D_{\tilde z}\tilde q&=\tilde\rho\partial_{\tilde\rho}\tilde q
 +\tilde z\partial_{\tilde z}\tilde q+\tilde q
 \  ,
 &\tilde D_{\varphi}\tilde\xi&=0 \ .  \label{eqtdtrhtq}
\end{align}
In terms of these operators equations \eqref{eqthdelxi2}-\eqref{eqthdelq2} can be written formally in the same structure as in \eqref{eqthdelxi}-\eqref{eqthdelq},
\begin{align}
 \Theta\tilde\Delta\tilde\xi=&2\left(
 \overline{\tilde\xi}\,\tilde D^{\tilde a}\tilde\xi
 -\overline{\tilde q}\,\tilde D^{\tilde a}\tilde q\right)
 \tilde D_{\tilde a}\tilde\xi \ , \label{eqthdelxi3} \\
 \Theta\tilde\Delta\tilde q=&2\left(
 \overline{\tilde\xi}\,\tilde D^{\tilde a}\tilde\xi
 -\overline{\tilde q}\,\tilde D^{\tilde a}\tilde q\right)
 \tilde D_{\tilde a}\tilde q \ , \label{eqthdelq3}
\end{align}
where the indices are raised and lowered using the metric $\tilde h_{\tilde a\tilde b}$. When writing out these equations in detail, the factors $e^{-2\gamma}$ appearing at both sides cancel, so they can also be considered as equations defined on a flat metric $\mathrm{diag}(1,1,\tilde\rho^2)$. These equations can be used to solve for $\tilde\xi$ and $\tilde q$ even if $\gamma$ is not known yet.

The Ricci tensor $\tilde R_{\tilde a\tilde b}$ belonging to the new unphysical metric $\tilde h_{\tilde a\tilde b}$ has the same form as in \eqref{eqricciab1},
\begin{equation}
 \tilde R_{\tilde a\tilde b}=\left(
 \begin{array}{ccc}
  -\partial^2_{\tilde\rho}\gamma
  +\frac{1}{\tilde\rho}\partial_{\tilde\rho}\gamma-\partial^2_{\tilde z}\gamma
  & \frac{1}{\rho}\partial_{\tilde z}\gamma & 0 \\
  \frac{1}{\tilde\rho}\partial_{\tilde z}\gamma
  & -\partial^2_{\tilde\rho}\gamma
  -\frac{1}{\tilde\rho}\partial_{\tilde\rho}\gamma-\partial^2_{\tilde z}\gamma & 0 \\
  0 & 0 & 0
 \end{array}
 \right) \ . \label{eqricciab2}
\end{equation}
However, the Einstein equation \eqref{eqthrrab} is only valid for the physical metric $h_{\tilde a\tilde b}$, with Ricci tensor components given in \eqref{eqrrtt11}-\eqref{eqrrtt22}. From the linear combination of the equations containing $R_{\tilde\rho\tilde\rho}+R_{\tilde z\tilde z}$ we can express $\partial^2_{\tilde\rho}\gamma+\partial^2_{\tilde z}\gamma$ in terms of $\xi$ and $q$. Similarly, using the equations containing $R_{\tilde\rho\tilde\rho}-R_{\tilde z\tilde z}$ and $R_{\tilde\rho\tilde z}$ we can solve for the first derivatives $\partial_{\tilde\rho}\gamma$ and $\partial_{\tilde z}\gamma$. Although the resulting expressions are rather long in terms of $\tilde\xi$ and $\tilde q$, using the operator $\tilde D_{\tilde a}$ given in \eqref{eqtdtrhtxi}-\eqref{eqtdtrhtq} and defining
\begin{equation}
 \tilde s_{\tilde a}=\tilde r\left(\tilde\xi\tilde D_{\tilde a}\tilde q
 -\tilde q\tilde D_{\tilde a}\tilde\xi\right) \label{eqstildedef}
\end{equation}
as in \cite{Sotiriou04}, the field equation containing the Ricci tensor can be written into the form
\begin{equation}
 \Theta^2\tilde R_{\tilde a\tilde b}=2\,\mathrm{Re}\left(
 \tilde D_{\tilde a}\tilde\xi\tilde D_{\tilde b}\overline{\tilde\xi}
 -\tilde D_{\tilde a}\tilde q\tilde D_{\tilde b}\overline{\tilde q}
 +\tilde s_{\tilde a}\overline{\tilde s}_{\tilde b}
 \right) \ , \label{eqth2rab1}
\end{equation}
which has formally the same structure as the original equation \eqref{eqthrrab}. We note that in \cite{Hoenselaers90} the factor $\tilde r$ was missed in the definition \eqref{eqstildedef} of $\tilde s_{\tilde a}$, which caused some errors in the final expressions of the multipole moments.

\subsection{Multipole moments}

As we have seen in Subsection \ref{secgenmultipole}, the choice $\phi=\xi$ gives the gravitational moment tensors $P^{(n)}_{\tilde a_1\ldots\tilde a_{n}}$, while $\phi=q$ yields the electromagnetic moments $Q^{(n)}_{\tilde a_1\ldots\tilde a_{n}}$. We have also introduced the unifying notation $M^{(n)}_{\tilde a_1\ldots\tilde a_{n}}$, which can be either of these two tensors. The conformally transformed $\tilde\phi=\tilde\Omega^{-1/2}\phi$, which is used to start the recursion in \eqref{eqpnp1ind0}-\eqref{eqpnp1ind}, is given in \eqref{eqhatxihatq}.

In the axially symmetric case each $M^{(n)}_{\tilde a_1\ldots\tilde a_{n}}$ multipole moment tensor is necessarily proportional to the tensor $\mathscr{C}(n_{\tilde a_1}\ldots n_{\tilde a_n})$, where $n^{\tilde a}$ is the unit vector at $\Lambda$ parallel to the rotational axis. It follows that each moment tensor is determined by a single scalar, i.e.
\begin{equation}
 M^{(n)}_{\tilde a_1\ldots\tilde a_{n}}=
 \hat{M}_n\,\mathscr{C}(n_{\tilde a_1}\ldots n_{\tilde a_n})\bigl|_{\Lambda} \label{eqpnaala}
\end{equation}
for some constants $\hat{M}_n$. Since at the point $\Lambda$ necessarily $\gamma=0$, the components of the vector are $n^{\tilde a}=(0,1,0)$ and $n_{\tilde a}=(0,1,0)$. The scalar moments for axial symmetry are defined as \cite{Hansen74}
\begin{equation}
 M_n=\frac{1}{n!}M^{(n)}_{\tilde a_1\ldots\tilde a_{n}}
 n^{\tilde a_1}\ldots n^{\tilde a_n}\bigl|_{\Lambda}\equiv
 \frac{1}{n!}M^{(n)}_{\tilde z\ldots\tilde z}
  \ . \label{eqpneqfacpaa}
\end{equation}
In particular, the scalar gravitational moments are $P_n=P^{(n)}_{\tilde z\ldots\tilde z}/n!$, and the scalar electromagnetic moments are  $Q_n=Q^{(n)}_{\tilde z\ldots\tilde z}/n!$. The mass and electric charge of the system has to be real, and given by $P_0\equiv M$ and $Q_0\equiv Q$, respectively. Setting a center of mass reference system, the gravitational dipole moment is pure imaginary, and gives the angular momentum of the configuration by $P_1=i J$.

It can be shown that (see e.g.~Appendix of \cite{Fodor89})
\begin{equation}
 n^{\tilde a_1}\ldots n^{\tilde a_n}
 \,\mathscr{C}(n_{\tilde a_1}\ldots n_{\tilde a_n})\Bigl|_{\Lambda}=
 \frac{n!}{(2n-1)!!}=\frac{2^n(n!)^2}{(2n)!} \ .
\end{equation}
Hence by substituting \eqref{eqpnaala} into \eqref{eqpneqfacpaa} we obtain that
\begin{equation}
 M_n=\frac{1}{(2n-1)!!}\hat{M}_n
 =\frac{2^n n!}{(2n)!}\hat{M}_n \ . \label{eqpnhatpn}
\end{equation}
Some authors define the scalar moments as $\hat{M}_n$ instead of $M_n$ \cite{Backdahl05,Backdahl07}.

Changing the conformal factor as $\tilde\Omega=\tilde\omega\Omega$, the multipole moment tensors transform according to \eqref{eqbarmmna}. In the axially symmetric case only the axis directional component of $\left(\partial_{\tilde a}\tilde\omega\right)_{\Lambda}$ is nonzero, so the transformation depends just on a single number. This corresponds to a translation of the configuration along the axis direction. Using \eqref{eqpnaala}, \eqref{eqpnhatpn}, and that $\left(\partial_{\tilde a}\tilde\omega\right)_{\Lambda}$ is parallel to $n_{\tilde a}|_\Lambda$, we can obtain the following simple formula for the transformation of the scalar multipole moments \cite{Fodor91}:
\begin{equation}
 \tilde M_n=\sum_{k=0}^n\binom{n}{k}\left[
 -\frac{1}{2}\left(\partial_{\tilde z}\tilde\omega\right)_\Lambda
 \right]^{n-k} M_k \ . \label{eqscalarmtr}
\end{equation}

\subsection{Complex null vector}

We follow the method introduced by Bäckdahl and Herberthson \cite{Backdahl05,Backdahl07}. Using the coordinates $\tilde x^{\tilde a}=(\tilde\rho,\tilde z,\varphi)$ we define the complex vector
\begin{equation}
 \boldsymbol{\eta}=\frac{\partial}{\partial\tilde z}
 -i\frac{\partial}{\partial\tilde\rho} \ ,
\end{equation}
which has the components $\eta^{\tilde a}=(-i,1,0)$. From the metric form \eqref{eqhabaxisyminf} it is easy to see that it is a null vector, $\tilde h_{\tilde a\tilde b}\eta^{\tilde a}\eta^{\tilde b}=0$. For the covariant derivative in the direction $\boldsymbol{\eta}$ we use the notation $\tilde\nabla_{\boldsymbol{\eta}}=\eta^{\tilde a}\tilde\nabla_{\tilde a}$. Acting on scalars this agrees with the directional derivative $\partial_{\boldsymbol{\eta}}=\eta^{\tilde a}\partial_{\tilde a}$. The following important property can be checked by direct calculation:
\begin{equation}
 \tilde\nabla_{\boldsymbol{\eta}}\eta^{\tilde a}=
 2\eta^{\tilde a}\partial_{\boldsymbol{\eta}}
 \gamma \ . \label{hatetanabeta}
\end{equation}

If we multiply the recursion definition \eqref{eqpnp1ind} at all indices by $\eta^{\tilde a_i}$, we obviously do not have to symmetrize. Furthermore, since $\eta^{\tilde a_i}$ is a null vector, the terms that must be added to make the tensor trace-free do not contribute either. The use of the complex null vector $\eta^{\tilde a}$ completely eliminates the problem of taking the symmetric trace-free part. Introducing
\begin{equation}
 f_n=\eta^{\tilde a_1}\ldots\eta^{\tilde a_n}
 \mathcal{P}^{(n)}_{\tilde a_1\ldots\tilde a_{n}} \ ,
\end{equation}
the first term on the right-hand side of \eqref{eqpnp1ind} can be written as
\begin{align}
 \eta^{\tilde a_1}\ldots\eta^{\tilde a_{n+1}}
 \tilde\nabla_{\tilde a_{n+1}}\mathcal{P}^{(n)}_{\tilde a_1\ldots\tilde a_{n}}
 =&\tilde\nabla_{\boldsymbol{\eta}}f_n
 -\mathcal{P}^{(n)}_{\tilde a_1\ldots\tilde a_{n}}\tilde\nabla_{\boldsymbol{\eta}}
 \left(\eta^{\tilde a_1}\ldots\eta^{\tilde a_{n}}\right) \notag\\
 =&\partial_{\boldsymbol{\eta}}f_n
 -2n f_n\partial_{\boldsymbol{\eta}}\gamma \ ,
\end{align}
where \eqref{hatetanabeta} has been used to get the second line. This way, \eqref{eqpnp1ind0}-\eqref{eqpnp1ind} provide the recursion formula for $f_n$, which has been given first in \cite{Backdahl05},
\begin{align}
 f_{0}&=\tilde\phi \ , \label{eqfnind0}\\
 f_{1}&=\partial_{\boldsymbol{\eta}}f_0 \ , \label{eqfnind1}\\
 f_{n+1}&=\partial_{\boldsymbol{\eta}}f_n
 -2n f_n\partial_{\boldsymbol{\eta}}\gamma
 -\frac{1}{2}n(2n-1)\eta^{\tilde a}
 \eta^{\tilde b}\tilde R_{\tilde a\tilde b}f_{n-1} \ . \label{eqfnind2}
\end{align}
We note that our function $\gamma$ has been denoted by $\beta$ in \cite{Backdahl05,Backdahl07}. In those papers there is also a further conformal transformation, specified by a function $\kappa$. Our choice \eqref{eqhabaxisyminf} of the 3-metric corresponds to $\kappa=\beta$ there.

Multiplying \eqref{eqpnaala} at all indices by $\eta^{\tilde a_i}$, since $\eta^{\tilde a}n_{\tilde a}=1$ we get $\hat M_n=f_n\bigr|_\Lambda$. From \eqref{eqpnhatpn} it can be seen that the scalar moments can be obtained by calculating the values of the functions $f_n$ at conformal infinity,
\begin{equation}
 M_n=\frac{1}{(2n-1)!!}f_n\bigr|_\Lambda
 =\frac{2^n n!}{(2n)!} \,f_n\bigr|_\Lambda \ . \label{eqmcpnfn}
\end{equation}

Using the form \eqref{eqricciab2} of the Ricci tensor, it is easy to see that
\begin{equation}
 \eta^{\tilde a}\eta^{\tilde b}\tilde R_{\tilde a\tilde b}=
 -\frac{2i}{\rho}\partial_{\boldsymbol{\eta}}\gamma \ . \label{eqetaetarab1}
\end{equation}
On the other hand, from \eqref{eqth2rab1} we obtain
\begin{equation}
 \Theta^2\eta^{\tilde a}\eta^{\tilde b}\tilde R_{\tilde a\tilde b}=
 2\eta^{\tilde a}\eta^{\tilde b}\left(
 \tilde D_{\tilde a}\tilde\xi\tilde D_{\tilde b}\overline{\tilde\xi}
 -\tilde D_{\tilde a}\tilde q\tilde D_{\tilde b}\overline{\tilde q}
 +\tilde s_{\tilde a}\overline{\tilde s}_{\tilde b}
 \right) \ , \label{eqthetetr}
\end{equation}
where the action of the operator $\tilde D_{\tilde a}$ has been given in \eqref{eqtdtrhtxi}-\eqref{eqtdtrhtq}.

\subsection{Leading order function}

Since the tensors $\mathcal{P}^{(n)}_{\tilde a_1\ldots\tilde a_{n}}$ are symmetric in their indices, they have in general $(n+1)(n+2)/2$ independent components. Adding trace terms, the quantities $S^{(n)}_{\tilde a}$ for $0\leq\tilde a\leq n$ have been introduced in \cite{Fodor89}, which decreased the necessary components to $n+1$. The introduction of the functions $f_n$ in \cite{Backdahl05} reduces the number of components to one at each order $n$, which significantly simplifies the calculation of the moments.

A further big simplification arises from the fact that all derivatives are taken in the $\boldsymbol{\eta}$ direction, and hence in the appropriate sense we can consider $f_n$ as functions of a single variable instead of functions of $\tilde\rho$ and $\tilde z$. In order to make use of this idea, the concept of leading order functions has been introduced in \cite{Backdahl05,Backdahl07}. The naming comes from the leading order terms of Legendre polynomials used initially for the static case in \cite{Herberthson04,Backdahl05a}.

Let us assume that $f$ is a complex valued axially symmetric analytic function in a neighborhood of $\Lambda$. Then it can also be considered as an analytic function $f(\tilde\rho,\tilde z)$ on the plane $\mathbb{R}^2$, satisfying $f(-\tilde\rho,\tilde z)=f(\tilde\rho,\tilde z)$. We extend the function $f$ into the complex $\tilde\rho$ plane by using the expansion of it in powers of $\tilde\rho$ and $\tilde z$. The \emph{leading order part}  of the function $f(\tilde\rho,\tilde z)$ is a function depending on a real parameter $\zeta$, which is defined as
\begin{equation}
 f_L(\zeta)=f(-i\zeta,\zeta) \ . \label{eqfldef}
\end{equation}
In other words, we substitute $\tilde z=\zeta$ and $\tilde \rho=-i\zeta$. It can be easily checked that
\begin{equation}
 \left(\tilde\nabla_{\boldsymbol{\eta}}f\right)_L(\zeta)
 =f_L{'}(\zeta) \ ,
\end{equation}
where the prime denotes differentiation with respect to $\zeta$. We do not write out the argument $(\zeta)$ of the leading order parts from now on.

If we write the expansion of $f$ in the form
\begin{equation}
 f(\tilde\rho,\tilde z)
 =\sum_{k=0}^\infty\sum_{l=0}^\infty a_{kl}\tilde\rho^k\tilde z^l
 =\sum_{N=0}^\infty\sum_{k=0}^{N}a_{k,N-k}\tilde\rho^k\tilde z^{N-k}
 \ , \label{eqfsumsum}
\end{equation}
where $a_{kl}$ are complex constants, then the leading order part function is
\begin{equation}
 f_L=\sum_{N=0}^\infty\tilde a_N\zeta^N \ \ , \ \ \ 
 \tilde a_N=\sum_{k=0}^{N}a_{k,N-k}(-i)^k \ . \label{eqflanl}
\end{equation}
If $f$ represents an axially symmetric function on the spacetime which is regular on the axis, then $f(-\tilde\rho,\tilde z)=f(\tilde\rho,\tilde z)$, and $a_{kl}=0$ for odd $k$. For this type of functions complex conjugation has the property $\left(\,\overline{f}\,\right)_L=\overline{(f_L)}$. In this case we also have
\begin{equation}
 \overline{\left(\tilde\nabla_{\boldsymbol{\eta}}f\right)_L}=
 \left(\tilde\nabla_{\boldsymbol{\eta}}\overline{f}\right)_L \ \ , \ \ \
 \overline{f_L{'}}=\overline{f}_L{'} \ .
\end{equation}
This also shows that the leading order part of regular real functions is real, in particular $\gamma_L$ and $\gamma_L{'}$ are real.

The leading order part of sums, products or quotients of functions is equal to the sum, product or quotient of the leading order parts, respectively. Introducing the notation
\begin{equation}
 \tilde R_L=\left(\eta^{\tilde a}\eta^{\tilde b}
 \tilde R_{\tilde a\tilde b}\right)_L \ , \label{eqhatrl}
\end{equation}
from \eqref{eqetaetarab1} we get
\begin{equation}
 \gamma_L{'}=\frac{\zeta}{2}\tilde R_L \ . \label{eqgammalpr}
\end{equation}
This shows that $\tilde R_L$ also has to be real. Since the leading order part of $\tilde r^2$ is zero, and $\Theta=\tilde r^2\tilde\xi\overline{\tilde\xi}-\tilde r^2\tilde q\overline{\tilde q}-1$, it follows that $\Theta_L=-1$. In order to calculate the leading order part of \eqref{eqthetetr} we can first check that
\begin{equation}
 \left(\eta^{\tilde a}\tilde D_{\tilde a}f\right)_L=
 2\zeta f_L{'}+f_L
 =2\sqrt{\zeta}\left(\sqrt{\zeta}\, f_L\right)'\ .
\end{equation}
Furthermore, since there is a factor $\tilde r$ in the definition \eqref{eqstildedef} of $\tilde s_{\tilde a}$, it follows that $\left(\eta^{\tilde a}\tilde s_{\tilde a}\right)_L=0$. Hence, the leading order part of \eqref{eqthetetr} can be written as
\begin{equation}
 \tilde R_L=2\bigl|2\zeta\tilde\xi_L{'}+\tilde\xi_L\bigr|^2
 -2\bigl|2\zeta\tilde q_L{'}+\tilde q_L\bigr|^2 \ . \label{eqrrllxiq}
\end{equation}

As we have seen in \eqref{eqmcpnfn}, the multipole moments are given by the values of the functions $f_n$ at the point $\Lambda$. To simplify the appearance of the expressions we introduce the notation
\begin{equation}
 y_n=\left(f_n\right)_L \ ,
\end{equation}
where $y_n$ are functions of $\zeta$. Then the scalar moments for the axially symmetric case can be calculated as
\begin{equation}
 M_n=\frac{1}{(2n-1)!!}y_n\bigr|_{\zeta=0}
 =\frac{2^n n!}{(2n)!} \,y_n\bigr|_{\zeta=0} \ . \label{eqpnynfac}
\end{equation}

From \eqref{eqfnind0}-\eqref{eqfnind2} we get the recursive definition of $y_n$ \cite{Backdahl05},
\begin{align}
 y_{0}&=\tilde\phi_L \ , \label{eqynind0}\\
 y_{1}&=y_0{'} \ , \label{eqynind1}\\
 y_{n+1}&=y_n{'}
 -2n\, y_n\gamma_L{'}
 -\frac{1}{2}n(2n-1)
 \tilde R_L\,y_{n-1} \ . \label{eqynind2}
\end{align}
If $\tilde\xi$ and $\tilde q$ is known, then $\tilde\xi_L$ and $\tilde q_L$ corresponding to $\tilde\phi_L$ can be calculated easily by \eqref{eqfldef} or \eqref{eqflanl}. Then $\tilde R_L$ is given by  \eqref{eqrrllxiq}, and $\gamma_L{'}$ by \eqref{eqgammalpr}. According to \eqref{eqpnynfac}, the gravitational moments $P_n\equiv M_n$ are obtained from the choice $\tilde\phi=\tilde\xi$, and the electromagnetic moments $Q_n\equiv M_n$ from $\tilde\phi=\tilde q$.

\section{Expansion and multipole moments} \label{secexpmulti}

\subsection{Expansion along the axis} \label{secexpaxis}

The conformally rescaled potentials $\tilde\xi$ and $\tilde q$ has been defined in \eqref{eqhatxihatq}. If the value of these potentials on the rotation axis is known, then the Ernst equations determine them on the whole space. Because of the asymptotic flatness  the functions $\tilde\xi$ and $\tilde q$ must be smooth at the point $\Lambda$. Hence we specify their axis values by the series expansion coefficients $m_n$ and $q_n$,
\begin{equation}
 \tilde\xi=\sum_{n=0}^{\infty}m_{n}\tilde z^n
 \ \ , \ \ \
 \tilde q=\sum_{n=0}^{\infty}q_{n}\tilde z^n \ .
\end{equation}

According to \eqref{eqrhotilztil}, on the axis $\tilde z=1/z$, where $z$ is the axial coordinate in the Weyl-Lewis-Papapetrou form \eqref{eqdsweyl} of the metric. Since we use the conformal factor $\Omega=\tilde r^2$, and along the axis $\tilde r=|\tilde z|=1/|z|$, it follows from \eqref{eqhatxihatq} that the expansions of the original, non transformed Ernst potentials on the axis are
\begin{equation}
 \xi=\frac{1}{|z|}\sum_{n=0}^{\infty}\frac{m_{n}}{z^{n}}
 \ \ , \ \ \
 q=\frac{1}{|z|}\sum_{n=0}^{\infty}\frac{q_{n}}{z^{n}} \ . \label{eqxiqaxisexp}
\end{equation}
This shows that the $1/z$ expansions of these functions necessarily differ in a minus sign on the upper and lower part of the rotational axis.

We note that this definition of the coefficients $m_n$ and $q_n$ is not as coordinate system specific as it may appear at first sight. In \eqref{eqxiqaxisexp} the coordinate $z$ acts as a parametrization along the line representing the symmetry axis, and its value is unimportant elsewhere. This parametrization satisfies $h_{ab}\left(\frac{\partial}{\partial z}\right)^a\left(\frac{\partial}{\partial z}\right)^b=1$ on the axis where $\gamma=0$, and $h_{ab}$ is given by \eqref{eqhabaxisym}. Generally, it is easy to find such parametrization even if the metric is not given in the Weyl-Lewis-Papapetrou form.

However, a freedom of constant shift remains in $z$. Setting $z=\hat z-z_0$, where $z_0$ is a constant, the potential $\xi$ can be also expanded in terms of $\hat z$,
\begin{equation}
 \xi=\frac{1}{|\hat z|}\sum_{n=0}^{\infty}\frac{\hat m_{n}}{\hat z^{n}} \ \ , \quad
 q=\frac{1}{|\hat z|}\sum_{n=0}^{\infty}\frac{\hat q_{n}}{\hat z^{n}} \ ,
\end{equation}
where
\begin{equation}
 \hat m_n=\sum_{k=0}^n\binom{n}{k}z_0^{n-k} m_k \ \ , \quad
 \hat q_n=\sum_{k=0}^n\binom{n}{k}z_0^{n-k} q_k
 \ . \label{eqexpmtr}
\end{equation}
This shows that the coefficients $m_n$ and $q_n$ transform in the same way as the scalar multipole moments in \eqref{eqscalarmtr}. In order to make $m_n$ and $q_n$ unique, we use this transformation to make real part of the gravitational dipole moment $m_1$ zero, choosing a center of mass system.

\subsection{Expansion of the Ernst equations}

We look for the solution of the Ernst equations \eqref{eqthdelxi3}-\eqref{eqthdelq3} in the power series form
\begin{equation}
 \tilde\xi=\sum_{\substack{k=0\\l=0}}^{\infty}a_{kl}\tilde\rho^k\tilde z^l
 \ \ , \ \ \
 \tilde q=\sum_{\substack{k=0\\l=0}}^{\infty}b_{kl}\tilde\rho^k\tilde z^l \ .
\end{equation}
The potentials $\tilde\xi$ and $\tilde q$ have to be smooth and regular on the rotation axis, which implies that the expansion contains only even powers of $\tilde\rho$. Hence for odd $k$ necessarily $a_{kl}=b_{kl}=0$. Obviously,
\begin{equation}
 a_{0l}=m_l \ \ , \ \ \ b_{0l}=q_l \ .
\end{equation}
We intend to calculate the coefficients $a_{kl}$ and $b_{kl}$ in terms of the expansion coefficients $m_l$ and $q_l$.

Since there have been a mistake in the equations for $a_{kl}$ and $b_{kl}$ published in the literature, we give a more detailed presentation here. Expanding out equation \eqref{eqthdelxi3}, we keep the linear terms on the left-hand side, and on the other side in curly brackets we group those terms together which have the same behavior in powers of $\tilde\rho$ and $\tilde z$,
\begin{align}
 \tilde\xi_{,\tilde\rho\tilde\rho}
 &+\frac{1}{\tilde\rho}\tilde\xi_{,\tilde\rho}
 +\tilde\xi_{,\tilde z\tilde z}
 =\biggl\{
 \left(\tilde\xi\overline{\tilde\xi}-\tilde q\overline{\tilde q}\right)
 \left(\tilde\rho^2\tilde\xi_{,\tilde\rho\tilde\rho}
 +\tilde\rho\tilde\xi_{,\tilde\rho}
 +\tilde z^2\tilde\xi_{,\tilde z\tilde z}
 \right) \notag\\
 &+2\tilde\rho\tilde z\tilde\xi_{,\tilde z}
 \left(\overline{\tilde\xi}\tilde\xi_{,\tilde\rho}
 -\overline{\tilde q}\tilde q_{,\tilde\rho}\right)
 +2\tilde\rho\tilde z\tilde\xi_{,\tilde\rho}
 \left(\overline{\tilde\xi}\tilde\xi_{,\tilde z}
 -\overline{\tilde q}\tilde q_{,\tilde z}\right) \notag\\
 &-2\left(\tilde\rho\tilde\xi_{,\tilde\rho}
 +\tilde z\tilde\xi_{,\tilde z}+\tilde\xi\right)
 \left[\,
 \overline{\tilde\xi}
 \left(\tilde\rho\tilde\xi_{,\tilde\rho}
 +\tilde z\tilde\xi_{,\tilde z}+\tilde\xi\right)
 -\overline{\tilde q}
 \left(\tilde\rho\tilde q_{,\tilde\rho}
 +\tilde z\tilde q_{,\tilde z}+\tilde q\right)
 \right]\biggr\} \notag\\
 &+\left\{
 \tilde z^2\left(\tilde\xi\overline{\tilde\xi}-\tilde q\overline{\tilde q}\right)
 \left(\tilde\xi_{,\tilde\rho\tilde\rho}
 +\frac{1}{\tilde\rho}\tilde\xi_{,\tilde\rho}\right)
 -2\tilde z^2\tilde\xi_{,\tilde\rho}
 \left(\overline{\tilde\xi}\tilde\xi_{,\tilde\rho}
 -\overline{\tilde q}\tilde q_{,\tilde\rho}\right)
 \right\}  \label{eqxirhorhopl}\\
 &+\left\{
 \tilde\rho^2\left(\tilde\xi\overline{\tilde\xi}-\tilde q\overline{\tilde q}\right)
 \tilde\xi_{,\tilde z\tilde z}
 -2\tilde\rho^2\tilde\xi_{,\tilde z}
 \left(\overline{\tilde\xi}\tilde\xi_{,\tilde z}
 -\overline{\tilde q}\tilde q_{,\tilde z}\right)
 \right\} \ . \notag
\end{align}
Substituting the expansions of $\tilde\xi$ and $\tilde q$, the right-hand side is a sum of terms
\begin{align}
 &\left(a_{kl}\overline{a}_{mn}-b_{kl}\overline{b}_{mn}\right)a_{pq}
 \,\times \label{eqrhsexp}\\
 &\ \ \times\Bigl\{
 \left(p^2+q^2-2pk-2ql-2p-3q-2k-2l-2\right)
 \tilde\rho^{k+m+p}\tilde z^{l+n+q} \notag\\
 &\ \ \ \ \ \ +p(p-2k)\tilde\rho^{k+m+p-2}\tilde z^{l+n+q+2}
 +q(q-1-2l)\tilde\rho^{k+m+p+2}\tilde z^{l+n+q-2}
 \Bigr\} \notag
\end{align}
for all $k$, $l$, $m$, $n$, $p$, $q$ nonnegative integers. The three kind of terms correspond to the three curly brackets in \eqref{eqxirhorhopl}. The integer $q$ here is obviously different from the Ernst potential $q$ in \eqref{eqxiqaxisexp}, however, we keep this notation to make comparison with earlier papers easier. Equating the $\tilde\rho^r\tilde z^s$ terms for nonnegative integers $r$ and $s$, we obtain the recursion relation for the components of $\tilde\xi$,
\begin{align}
 (r+2)^2& a_{r+2,s}=-(s+2)(s+1)a_{r,s+2} \notag\\
 &+\sum_{\substack{k+m+p=r\\l+n+q=s}}
 \left(a_{kl}\overline{a}_{mn}-b_{kl}\overline{b}_{mn}\right) \times \label{eqarsrec} \\
 &\ \ \times\Bigl[a_{pq}\left(p^2+q^2-2p-3q-2k-2l-2pk-2ql-2\right) \notag\\
 &\ \ \ \ +a_{p+2,q-2}(p+2)(p+2-2k)
 +a_{p-2,q+2}(q+2)(q+1-2l)\Bigr] \ . \notag
\end{align}
Equation \eqref{eqthdelq3} can be obtained from \eqref{eqthdelxi3} by exchanging $\tilde\xi$ and $\tilde q$, followed by reversing the signature of the cubic terms. The recursion for the components of $\tilde q$ can be written as
\begin{align}
 (r+2)^2& b_{r+2,s}=-(s+2)(s+1)b_{r,s+2} \notag\\
 &+\sum_{\substack{k+m+p=r\\l+n+q=s}}
 \left(a_{kl}\overline{a}_{mn}-b_{kl}\overline{b}_{mn}\right) \times \label{eqbrsrec} \\
 &\ \ \times\Bigl[b_{pq}\left(p^2+q^2-2p-3q-2k-2l-2pk-2ql-2\right) \notag\\
 &\ \ \ \ +b_{p+2,q-2}(p+2)(p+2-2k)
 +b_{p-2,q+2}(q+2)(q+1-2l)\Bigr] \ . \notag
\end{align}

In papers \cite{Hoenselaers90,Sotiriou04} in the third line of the equations corresponding to \eqref{eqarsrec}-\eqref{eqbrsrec} we can find $-4p-5q$ instead of the correct $-2p-3q-2k-2l$ terms. The mistake has been corrected in the proceedings paper \cite{Perjes03}, but it may have remained unnoticed. For the purely gravitational case, when $b_{kl}=0$, the two expressions are equivalent because of the symmetry of the $a_{kl}a_{pq}$ product. This shows that the corresponding expression in \cite{Fodor89} is still correct. Calculating $a_{rs}$ and $b_{rs}$ up to some order in $r+s$, forming $\tilde\xi$ and $\tilde q$, then substituting back to \eqref{eqthdelxi3}-\eqref{eqthdelq3}, we have checked that the equations are satisfied up to order $r+s-2$. The same procedure with the originally published recursion formula fails, showing that the two expressions are clearly not equivalent. As we will see, the values of the multipole moments $P_n$ and $Q_n$ expressed in terms of $m_l$ and $q_l$ will also be influenced.

The sums in \eqref{eqarsrec}-\eqref{eqbrsrec} are essentially sums for four integer variables instead of six, since two integers are determined by the given value of $s$ and $r$. We have to be careful with the start of the summations in $p$ and $q$, because of the shift by $\pm 2$ of the indices in the second and third term of \eqref{eqrhsexp}. A natural way to calculate the sums in \eqref{eqarsrec}-\eqref{eqbrsrec} is
\begin{align}
 (r+2)^2& a_{r+2,s}=-(s+2)(s+1)a_{r,s+2} \notag\\
 &+\sum_{\substack{p=0\\ \mathrm{even}}}^r\sum_{q=0}^s
 \sum_{\substack{k=0\\ \mathrm{even}}}^{r-p}\sum_{l=0}^{s-q}
 \left(a_{kl}\overline{a}_{mn}-b_{kl}\overline{b}_{mn}\right) \times  \notag\\
 &\ \ \ \ \ \ \  \times a_{pq}\left(p^2+q^2-2p-3q-2k-2l-2pk-2ql-2\right)
 \label{eqarsrec2}\\
 &+\sum_{\substack{p=-2\\ \mathrm{even}}}^r\sum_{q=2}^s
 \sum_{\substack{k=0\\ \mathrm{even}}}^{r-p}\sum_{l=0}^{s-q}
 \left(a_{kl}\overline{a}_{mn}-b_{kl}\overline{b}_{mn}\right)
 a_{p+2,q-2}(p+2)(p+2-2k) \notag\\
 &+\sum_{\substack{p=2\\ \mathrm{even}}}^r\sum_{q=-2}^s
 \sum_{\substack{k=0\\ \mathrm{even}}}^{r-p}\sum_{l=0}^{s-q}
 \left(a_{kl}\overline{a}_{mn}-b_{kl}\overline{b}_{mn}\right)
 a_{p-2,q+2}(q+2)(q+1-2l) \ , \notag
\end{align}
\begin{align}
 (r+2)^2& b_{r+2,s}=-(s+2)(s+1)b_{r,s+2} \notag\\
 &+\sum_{\substack{p=0\\ \mathrm{even}}}^r\sum_{q=0}^s
 \sum_{\substack{k=0\\ \mathrm{even}}}^{r-p}\sum_{l=0}^{s-q}
 \left(a_{kl}\overline{a}_{mn}-b_{kl}\overline{b}_{mn}\right) \times  \notag\\
 &\ \ \ \ \ \ \  \times b_{pq}\left(p^2+q^2-2p-3q-2k-2l-2pk-2ql-2\right)
 \label{eqbrsrec2}\\
 &+\sum_{\substack{p=-2\\ \mathrm{even}}}^r\sum_{q=2}^s
 \sum_{\substack{k=0\\ \mathrm{even}}}^{r-p}\sum_{l=0}^{s-q}
 \left(a_{kl}\overline{a}_{mn}-b_{kl}\overline{b}_{mn}\right)
 b_{p+2,q-2}(p+2)(p+2-2k) \notag\\
 &+\sum_{\substack{p=2\\ \mathrm{even}}}^r\sum_{q=-2}^s
 \sum_{\substack{k=0\\ \mathrm{even}}}^{r-p}\sum_{l=0}^{s-q}
 \left(a_{kl}\overline{a}_{mn}-b_{kl}\overline{b}_{mn}\right)
 b_{p-2,q+2}(q+2)(q+1-2l) \ , \notag
\end{align}
where in each term we have to substitute $m=r-p-k$ and $n=s-q-l$. Obviously, $r$ can be assumed to be even, and we have to include only even values of $p$ and $k$ in the sums. Because of the $p+2$ factor, it would not be really necessary to include the terms with $p=-2$ in the second sum. However, the third sum would give an incorrect value if we started the summation from $q=0$, since the terms corresponding to $q=-1$ and nonzero $l$ are nonvanishing.

Our aim with \eqref{eqarsrec2}-\eqref{eqbrsrec2} is to express all $a_{kl}$ and $b_{kl}$ coefficients in terms of $a_{0l}=m_l$ and $b_{0l}=q_l$. While the two linear terms contain coefficients with $k+l=r+s+2$, the cubic summation terms only have coefficients with $k+l\leq r+s$. Denoting $r+s=N$, we can proceed by increasing the order in $N$ one by one. First, for $N=0$ we calculate $a_{20}$, then for $N=1$ we get $a_{21}$. Continuing with $N=2$, first for $r=0$ we get $a_{22}$ and then for $r=2$ we obtain $a_{40}$. Proceeding further in this way, parallelly for both $a_{kl}$ and $b_{kl}$, we can express the new coefficients by those that has been already calculated earlier. As supplementary material, we attach a Mathematica and an essentially equivalent Maple file (named "moments-general"), where we give a concrete implementation for this series expansion solution procedure, and also validate its correctness by substituting back into the Ernst equations.

\subsection{Expressions for the scalar moments}

Using \eqref{eqfsumsum}-\eqref{eqflanl}, the expansion of the leading order functions can be calculated as
\begin{align}
 \tilde\xi_L=\sum_{N=0}^\infty\tilde a_N\zeta^N \ \ , \ \ \ 
 \tilde a_N=\sum_{\substack{k=0\\
 \mathrm{even}}}^{N}a_{k,N-k}(-1)^{k/2} \ , \label{eqxilan}\\
 \tilde q_L=\sum_{N=0}^\infty\tilde b_N\zeta^N \ \ , \ \ \ 
 \tilde b_N=\sum_{\substack{k=0\\
 \mathrm{even}}}^{N}b_{k,N-k}(-1)^{k/2} \ . \label{eqqlbn}
\end{align}
In order to calculate the multipole moments up to order $N_{\mathrm{max}}$, we need to obtain first the values of $a_{kl}$ and $b_{kl}$ up to order $k+l=N_{\mathrm{max}}$. Then by \eqref{eqxilan}-\eqref{eqqlbn} we can get the leading order parts $\tilde\xi_L$ and $\tilde q_L$, also up to order $N_{\mathrm{max}}$ in $\zeta$. After this, $\tilde R_L$ can be calculated by \eqref{eqrrllxiq}, and $\gamma_L{'}$ by \eqref{eqgammalpr}. Since they will be used only from the second multipole moment, it is enough to calculate them up to order $N_{\mathrm{max}}-2$. Then we can use the recursive formula \eqref{eqynind0}-\eqref{eqynind2} to calculate the functions $y_n$. For the electromagnetic case we need to calculate two set of moments, and hence two sets of $y_n$ functions for $n\leq N_{\mathrm{max}}$. The gravitational moments $P_n\equiv M_n$ are obtained by choosing $\tilde\phi=\tilde\xi$, and the electromagnetic moments $Q_n\equiv M_n$ by setting $\tilde\phi=\tilde q$. Each $y_n$ has to be calculated up to order $N_{\mathrm{max}}-n$ in the variable $\zeta$. The two sets of scalar moments can be obtained by taking the values at $\zeta=0$, according to \eqref{eqpnynfac}.

In order to make the final expressions for the multipole moments simpler, we use the following notations introduced in \cite{Fodor89,Hoenselaers90,Sotiriou04}:
\begin{align}
 M_{ij}&=m_i m_j-m_{i-1}m_{j+1} \ , &
 S_{ij}&=m_i q_j-m_{i-1}q_{j+1} \ , \label{eqmijsij}\\
 Q_{ij}&=q_i q_j-q_{i-1}q_{j+1} \ , &
 H_{ij}&=q_i m_j-q_{i-1}m_{j+1} \ , \label{eqqijhij}
\end{align}
for $i>j\geq 0$ integers. For the first seven gravitational moments we obtain the following results
\begin{align}
 P_0&=m_0 \ , \label{eqp0m0} \\
 P_1&=m_1 \ , \\
 P_2&=m_2 \ , \\
 P_3&=m_3+\frac{1}{5}\overline{q}_0 S_{10} \ , \\
 P_4&=m_4
 -\frac{1}{7}\overline{m}_0 M_{20}
 +\frac{3}{35}\overline{q}_1 S_{10}
 +\frac{1}{7}\overline{q}_0(3S_{20}-2H_{20}) \ , \\
 P_5&=m_5
 -\frac{1}{21}\overline{m}_1 M_{20}
 -\frac{1}{3}\overline{m}_0 M_{30}
 +\frac{1}{21}\overline{q}_2 S_{10}
 +\frac{1}{21}\overline{q}_1(4S_{20}-3H_{20}) \notag\\
 &+\frac{1}{21}\overline{q}_0\left(
 \overline{q}_0 q_0 S_{10}
 -\overline{m}_0 m_0 S_{10}
 +14S_{30}+13S_{21}-7H_{30}
 \right) \ , \\
 P_6&=m_6
 -\frac{5}{231}\overline{m}_2 M_{20}
 -\frac{4}{33}\overline{m}_1 M_{30}
 +\frac{1}{33}\overline{m}_0^2 m_0 M_{20}
 -\frac{1}{33}\overline{m}_0(18M_{40}+8M_{31}) \notag\\
 &+\frac{1}{33}\overline{q}_3 S_{10}
 +\frac{1}{231}\overline{q}_2(25S_{20}-20H_{20})
 +\frac{2}{231}\overline{q}_1(35S_{30}+37S_{21}-21H_{30}) \notag\\
 &-\frac{1}{1155}(37\overline{q}_1\overline{m}_0
  +13\overline{q}_0\overline{m}_1)m_0 S_{10}
 +\frac{1}{33}\overline{q}_0^2
  \left(5q_0 S_{20}-4m_0 Q_{20}+3q_1 S_{10}\right) \notag\\
 &+\frac{10}{231}\overline{q}_1\overline{q}_0 q_0 S_{10}
 +\frac{2}{33}\overline{q}_0\overline{m}_0
  \left(2m_0 H_{20}-3q_0 M_{20}-2m_1 S_{10}\right) \\
 &+\frac{1}{33}\overline{q}_0
  \left(30S_{40}+32S_{31}-24H_{31}-12H_{40}\right) \ . \notag
\end{align}
The electromagnetic moments are
\begin{align}
 Q_0&=q_0 \ , \\
 Q_1&=q_1 \ , \\
 Q_2&=q_2 \ , \\
 Q_3&=q_3-\frac{1}{5}\overline{m}_0 H_{10} \ , \\
 Q_4&=q_4
 +\frac{1}{7}\overline{q}_0 Q_{20}
 -\frac{3}{35}\overline{m}_1 H_{10}
 -\frac{1}{7}\overline{m}_0(3H_{20}-2S_{20}) \ , \\
 Q_5&=q_5
 +\frac{1}{21}\overline{q}_1 Q_{20}
 +\frac{1}{3}\overline{q}_0 Q_{30}
 -\frac{1}{21}\overline{m}_2 H_{10}
 -\frac{1}{21}\overline{m}_1(4H_{20}-3S_{20}) \notag\\
 &+\frac{1}{21}\overline{m}_0\left(
 \overline{m}_0 m_0 H_{10}
 -\overline{q}_0 q_0 H_{10}
 -14H_{30}-13H_{21}+7S_{30}
 \right) \ , \\
 Q_6&=q_6
 +\frac{5}{231}\overline{q}_2 Q_{20}
 +\frac{4}{33}\overline{q}_1 Q_{30}
 +\frac{1}{33}\overline{q}_0^2 q_0 Q_{20}
 +\frac{1}{33}\overline{q}_0(18Q_{40}+8Q_{31}) \notag\\
 &-\frac{1}{33}\overline{m}_3 H_{10}
 -\frac{1}{231}\overline{m}_2(25H_{20}-20S_{20})
 -\frac{2}{231}\overline{m}_1(35H_{30}+37H_{21}-21S_{30})\notag\\
 &-\frac{1}{1155}(37\overline{m}_1\overline{q}_0
  +13\overline{m}_0\overline{q}_1)q_0 H_{10}
 +\frac{1}{33}\overline{m}_0^2
  \left(5m_0 H_{20}-4q_0 M_{20}+3m_1 H_{10}\right) \notag\\
 &+\frac{10}{231}\overline{m}_1\overline{m}_0 m_0 H_{10}
 +\frac{2}{33}\overline{m}_0\overline{q}_0
  \left(2q_0 S_{20}-3m_0 Q_{20}-2q_1 H_{10}\right) \label{eqq6q6}\\
 &-\frac{1}{33}\overline{m}_0
  \left(30H_{40}+32H_{31}-24S_{31}-12S_{40}\right) \ . \notag
\end{align}
These expressions can be checked and higher order results can be obtained by the Mathematica or Maple file (named "moments-general") attached as supplementary material. 

For $n\geq 3$ the expressions for $P_n$ and $Q_n$ are clearly different from the results published earlier in the literature \cite{Hoenselaers90,Sotiriou04}. The reason for the difference from the more recent result is the application of an incorrect version of Eqs.~\eqref{eqarsrec} and \eqref{eqbrsrec}. The most striking difference from the earlier results is that in the correct expressions generally $P_3\not=p_3$ and $Q_3\not=q_3$ in the electromagnetic case. Although the multipole moments shown above have been calculated by the simpler method using the leading order functions, we have checked that the earlier method using the quantities $S^{(n)}_{\tilde a}$ introduced in \cite{Fodor89} lead to identical results.

The structure of the expressions for $P_n$ and $Q_n$ are very similar. In fact, it is easy to obtain the result for $Q_n$ by taking the expression for $P_n$ and make the exchanges $m_n\longleftrightarrow q_n$ and $\overline{m}_n\longleftrightarrow -\overline{q}_n$, with a minus sign in the conjugated quantities. This also implies that we have to exchange $M_{ij}\longleftrightarrow Q_{ij}$ and $S_{ij}\longleftrightarrow H_{ij}$. This property follows from an analogous formal symmetry of the Ernst equations \eqref{eqthdelxi}-\eqref{eqthdelq}, for the exchanges $\xi\longleftrightarrow q$ and $\overline{\xi}\longleftrightarrow -\overline{q}$.

\section{Multipole moments of exact solutions}\label{secmagtssol}

We are living in an era in which we have the opportunity to make precise astrophysical observations, and we hope to test theoretical models and assumptions we have in General Relativity. In order to best describe the collected data theoretically, we need to use suitable models.
Because of black hole uniqueness results, the Kerr spacetime is likely to describe the object in the center of our galaxy. However, the Kerr metric is not suitable for the description of the exterior region of rotating compact stars, because its quadrupole moment is too small.
Even if extremely precise numerical solutions exist for the description of rotating neutron stars, exact solutions describing the exterior region can still give valuable insights \cite{Berti04,Berti05,Pappas13}.
There are powerful methods to generate axially symmetric stationary vacuum and electrovacuum solutions.

In this last section of the paper we will consider a 5-parameter stationary axially symmetric exact solution published in 2000 by Manko et al.~\cite{Manko00b}. These are part of a wider class of two-soliton \cite{Manko95} and N-soliton solutions \cite{Manko95b}, having the special property that they can be written in terms of prolate or oblate spheroidal coordinates.
The 5 real parameters determine the mass, angular momentum, quadrupole moment, charge and magnetic dipole moment of the solution.
As special cases, this 5-parameter solution includes the Kerr-Newman and the charged $\delta=2$ Tomimatsu-Sato solutions.

The Tomimatsu-Sato solutions \cite{tomimatsu72,tomimatsu73} are families of stationary axially symmetric vacuum solutions labeled by a positive integer parameter $\delta$. The $\delta=1$ family corresponds to the Kerr solution. The solutions up to $\delta=4$ were presented in 1972 by Tomimatsu and Sato \cite{tomimatsu72,tomimatsu73}, while the expressions for general $\delta\geq 1$ integer have been given a few years later by Yamazaki \cite{Yamazaki77a,Yamazaki77b}. The electrically charged generalizations have been constructed by Ernst \cite{Ernst73} and Yamazaki \cite{Yamazaki78}. A further generalization of the $\delta=2$ solution by adding a magnetic dipole parameter has been published by Manko et al.~in 1998 \cite{Manko98,Manko99}. This solution has been generalized further in \cite{Manko00b} by adding a fifth parameter, allowing the specification of the quadrupole moment and the inclusion of the Kerr-Newman solution.

Tomimatsu and Sato have already calculated the quadrupole moment in their original paper \cite{tomimatsu73}. Higher multipoles in the weak gravity limit in the hyperextreme case have been considered in \cite{Kinnersley74,Tanabe76}. The results for the moments of the $\delta=2, 3, 4$ vacuum Tomimatsu-Sato solutions have been presented up to order $P_{12}$ in \cite{Fodor91}. In the following we mainly focus on the charged magnetized 5-parameter generalization of the Kerr and $\delta=2$ Tomimatsu-Sato solution published in \cite{Manko00b}. The mass-quadrupole and the magnetic dipole moment of this solution have been already calculated in \cite{Manko00b}.

\subsection{Prolate and oblate spheroidal coordinates}

The metric of the Tomimatsu-Sato solutions and their generalizations have been originally given using Weyl-Lewis-Papapetrou coordinates \eqref{eqdsweyl}, where the functions $f$, $\omega$ and $\gamma$ are given functions of the coordinates $\rho$ and $z$. These three functions have been presented in \cite{tomimatsu72,tomimatsu73,Manko00b} using prolate spheroidal coordinates $x$ and $y$, which are defined by
\begin{equation}
\rho=\kappa\sqrt{x^2-1}\sqrt{1-y^2} \quad , \qquad z=\kappa x y \ ,
\end{equation}
where $\kappa$ is a positive constant. The range of the coordinates is $x>1$ and $1\geq y\geq-1$. A simple form of the inverse relation is \cite{Kramer80}
\begin{equation}
x=\frac{r_{+}+r_{-}}{2\kappa} \ \ , \quad
y=\frac{r_{+}-r_{-}}{2\kappa} \ \ , \quad
r_\pm=\sqrt{\rho^2+(z\pm\kappa)^2} \ . \label{eqxyrpm}
\end{equation}
In order to avoid the appearance of the square roots in the metric it is advantageous to use the coordinates $x$ and $y$ directly. Then the metric of the Tomimatsu-Sato solutions and their electromagnetic generalizations can be written as
\begin{align}
\mathrm{d}s^2=&-f\left(\mathrm{d}t
-\omega\mathrm{d}\varphi\right)^2 \label{eqdsprol}\\
&+\frac{\kappa^2}{f}\left[e^{2\gamma}\left(x^2-y^2\right)\left(
\frac{\mathrm{d}x^2}{x^2-1}+\frac{\mathrm{d}y^2}{1-y^2}\right)
+\left(x^2-1\right)\left(1-y^2\right)\mathrm{d}\varphi^2\right] \ . \notag
\end{align}

The value of the constant $\kappa$ is fixed by the freely specifiable parameters of the solution. If the angular momentum, charge or dipole moment is large enough, the solution may become hyperextreme, and $\kappa^2$ turns to be negative. It has been already realized by Tomimatsu and Sato in \cite{tomimatsu73} that hyperextreme extension of their solutions can be obtained by the complex transformation $\kappa\to -i\kappa$, $x\to i x$. The same is true for the electromagnetic generalizations. This correspond to using oblate spheroidal coordinates
\begin{equation}
\rho=\kappa\sqrt{x^2+1}\sqrt{1-y^2} \quad , \qquad z=\kappa x y
\end{equation}
with real $\kappa>0$. The spacetime metric is then
\begin{align}
\mathrm{d}s^2=&-f\left(\mathrm{d}t
-\omega\mathrm{d}\varphi\right)^2 \label{eqdsobl}\\
&+\frac{\kappa^2}{f}\left[e^{2\gamma}\left(x^2+y^2\right)\left(
\frac{\mathrm{d}x^2}{x^2+1}+\frac{\mathrm{d}y^2}{1-y^2}\right)
+\left(x^2+1\right)\left(1-y^2\right)\mathrm{d}\varphi^2\right] \ . \notag
\end{align}

Hyperextreme Tomimatsu-Sato solutions have been already discussed in \cite{Kinnersley74,Tanabe76,Manko97,Manko13}. Although hyperextreme solutions always contain naked singularities, these solutions can still describe well the exterior region of rotating bodies. Actually, it is known that the exterior region of relatively small mass rotating objects often correspond to hyperextreme spacetimes. Defining the angular momentum per unit mass as $a=\frac{J}{M}$, in the absence of electromagnetic fields the configuration is hyperextreme if $\frac{a}{M}>1$. For example, for the Earth $\frac{a}{M}\approx 740$, and for a vinyl LP record spinning on a turntable $\frac{a}{M}\approx 10^{18}$ \ \cite{Rosquist08,Dietz85}. 
The exterior region of a rotating disk of dust can also be hyperextreme \cite{Kleinwachter11,Breithaupt15}.
For main sequence stars $\frac{a}{M}$ may be as large as $100$, which has to decrease to $\frac{a}{M}\ll 1$ if they evolve into neutron stars \cite{Felice82}.
The $\frac{a}{M}$ ratio can also be as large as $500$ for the bulge of spiral galaxies \cite{Bradley91}.

For the subextreme case the metric on the 3-manifold $\mathcal{M}$ is
\begin{equation}
h_{ab}=\left(
\begin{array}{ccc}
\kappa^2 e^{2\gamma}\,\displaystyle\frac{x^2-y^2}{x^2-1} & 0 & 0 \\
0 & \kappa^2 e^{2\gamma}\,\displaystyle\frac{x^2-y^2}{1-y^2} & 0 \\
0 & 0 & \kappa^2\left(x^2-1\right)\left(1-y^2\right)
\end{array}
\right) \ . \label{eqhabxy}
\end{equation}
The spatial metric for the hyperextreme case can be easily seen from \eqref{eqdsobl}. It can also be obtained by the substitution $\kappa\to -i\kappa$, $x\to i x$, taking also into account the change in $\mathrm{d}x^2$.

\subsection{Five-parameter solution}

The five-parameter solution introduced by Manko et al.~\cite{Manko00b} contain several interesting special cases, such as the (electro-)vacuum Tomimatsu-Sato solutions with $\delta=1$ and $\delta=2$. The solution depends on five parameters: $m$, $\hat q$, $a$, $\mu$ and $b$. In this paper we use hat on the constant $q$ of \cite{Manko00b} in order to distinguish it from the electromagnetic Ernst potential defined in \eqref{eqxiqdef}.
In the absence of NUT charges and magnetic monopoles, $m$ and $\hat q$ are real, and correspond to the mass and electric charge, respectively. By shifting to a center of mass system the mass dipole moment can be made zero. Then the parameter $a$ is real and gives the angular momentum per unit mass of the solution. 
The parameters $\mu$ and $b$ are related, but not equal, to the magnetic dipole moment and mass-quadrupole moment.

We introduce the constants $c$ and $\nu$ by
\begin{equation}
 c=a-b \quad , \qquad \mu=\hat q\,\nu \ ,
\end{equation}
and will use them at most places instead of $a$ and $\mu$ in the following. This way the structure of the final expressions for the electromagnetic multipole moments will become quite similar to the structure of the gravitational moments. This also makes the expressions shorter, since the constant $a$ appears mostly in the combination $a-b$. 

To simplify the equations, Manko et al.~have introduced the following combinations of the constants,
\begin{equation}
\delta=\frac{\hat q^2\nu^2-m^2b^2}{m^2-c^2-\hat q^2}\quad ,\qquad 
d=\frac{1}{4}\left(m^2-c^2-\hat q^2\right) \ . \label{eqdeld}
\end{equation}
The parameters $b$, $\mu$, and consequently $c=a-b$ may be complex, but the solution in $x$, $y$ spheroidal coordinates only exists if the combination $d+\delta$ is real.
If $d+\delta>0$ real, then the solution is subextreme, and the spacetime metric is described by \eqref{eqdsprol}. In the hyperextreme case $d+\delta<0$ real, and the metric is given by \eqref{eqdsobl}. In both cases, we have a solution of the Einstein equations if
\begin{equation}
\kappa=\sqrt{\left|d+\delta\right|} \ . \label{eqkappasqrt}
\end{equation}
As we will see, the assumption of reflection symmetry with respect to the equatorial plane imposes further restriction on the constants.

The complex Ernst potentials are given in \cite{Manko00b} by the expressions
\begin{equation}
\mathcal{E}=\frac{A-2 m B}{A+2 m B} \quad , \qquad
\Phi=\frac{2C}{A+2 m B} \ , \label{eqephicmts}
\end{equation}
where in the subextreme case
\begin{align}
A=&4\left[\left(\kappa^2 x^2-\delta y^2\right)^2-d^2
-i\kappa^3 x y c\left(x^2-1\right)\right] \\  
&-\left(1-y^2\right)\left[c(d-\delta)-m^2b+\hat q^2\nu\right]
\left[c\left(y^2+1\right)+4 i\kappa x y\right]  \ , \notag \\
B=&\kappa x\left[2\kappa^2\left(x^2-1\right)
+\left(bc+2\delta\right)\left(1-y^2\right)\right] \\
&+iy\left[2\kappa^2 b \left(x^2-1\right)
-\left(\kappa^2 c-m^2 b+\hat q^2\nu-2a\delta\right)
\left(1-y^2\right)\right] \ , \notag\\
C=&2\kappa^2\hat q\left(x^2-1\right)(\kappa x+i\nu y) \label{eqccxy}\\
&+\hat q\left(1-y^2\right)\left\{ \kappa x\left(2\delta+\nu c\right)
-i y\left[c (d-\delta )-m^2 b+\hat q^2\nu-2\nu\delta\right]\right\} \ . \notag
\end{align}

The expressions for $A$, $B$ and $C$ in the hyperextreme case can be obtained by the substitution $\kappa\to -i\kappa$, $x\to i x$. It can be checked by a calculation, which is lengthy even for an algebraic manipulation software, that the Ernst potentials $\mathcal{E}$ and $\Phi$ really satisfy the Ernst equations \eqref{eqfdeleps}-\eqref{eqfdelphi}. The function $e^{2\gamma}$ drops out from these equations when the 3-metric is of the form \eqref{eqhabxy}.

\subsection{Axis expansion}

Along the upper part of the rotation axis $y=1$, and the axial Weyl coordinate can be given there by $z=\kappa x$. The choice of the factor $\kappa$ ensures that $z$ is a parametrization satisfying $h_{ab}\left(\frac{\partial}{\partial z}\right)^a\left(\frac{\partial}{\partial z}\right)^b=1$, as required in Subsection \ref{secexpaxis}. The complex potentials $\xi$ and $q$ are defined according to \eqref{eqxiqdef}. To shorten the resulting expressions we introduce one more notation for the following combination of the constants:
\begin{equation}
v=d-\delta \ . \label{eqvconstdef}
\end{equation}
For both the subextreme and the hyperextreme case, on the upper part of the axis we obtain
\begin{align}
\xi&=\frac{m(z+i b)}
{z^2-icz+v} \ , \label{eqxiaxis}\\
q&=\frac{\hat q(z+i \nu)}
{z^2-icz+v}  \ . \label{eqqiaxis}
\end{align}
According to \eqref{eqxiqaxisexp}, the first few expansion coefficients turn out to be
\begin{align}
m_0&=m \ , \label{eqm0cmts}\\
m_1&=im (c+b) \ , \\
m_2&=-m(c^2+v+bc) \ , \\
m_3&=-im\left[c^3+2cv+b\left(c^2+v\right)\right] \ , \\
m_4&=m\left[c^4+3c^2 v+v^2+b\left(c^3+2cv\right)\right] \ , \\
m_5&=im\left[c^5+4c^3 v+3cv^2+b\left(c^4+3c^2 v+v^2\right)\right] \ ,
\end{align}
and
\begin{align}
q_0&= \hat{q} \ , \\
q_1&=i \hat{q}(c+\nu) \ , \label{eqq1hatq}\\
q_2&=-\hat{q}(c^2+v+\nu c) \ , \\
q_3&=-i\hat{q}\left[c^3+2cv+\nu\left(c^2+v\right)\right]\ , \\
q_4&=\hat{q}\left[c^4+3c^2 v+v^2+\nu\left(c^3+2cv\right)\right] \ , \\
q_5&=i\hat{q}\left[c^5+4c^3 v+3cv^2+\nu\left(c^4+3c^2 v+v^2\right)\right]
\ . \label{eqq4cmts}
\end{align}
It is possible to give the general expression for the coefficients. Defining
\begin{eqnarray}
&s=\sqrt{c^2+4 v}\ , \\
&f_k=\dfrac{i^{k}}{2^{k+1}s}\left[(c+s)^{k+1}-(c-s)^{k+1}\right] \ ,
\end{eqnarray}
for all $k\geq0$ integers we have
\begin{equation}
m_k=m\left(f_{k}+ibf_{k-1}\right) \quad , \qquad
q_k=\hat{q}\left(f_{k}+i\nu f_{k-1}\right) \ . \label{eqmkqkf}
\end{equation}

It can already be seen from \eqref{eqxiaxis}-\eqref{eqqiaxis}, that if we make the replacements $m\leftrightarrow\hat{q}$, $b\leftrightarrow\nu$, keeping $c$ and $v$ fixed, then we obtain $q_n$ from $m_n$ and vice versa. However, this is just a formal symmetry, since if we write out $c$ and $v$ in detail, they are not invariant under this replacement.

\subsection{Lower order moments}

At this stage we can use \eqref{eqp0m0}-\eqref{eqq6q6} to calculate the multipole moments $P_n$ and $Q_n$ up to order $6$. Clearly $P_n=m_n$ and $Q_n=q_n$ only for $n=0, 1, 2$. It can be seen that for the 5-parameter generalization of the Tomimatsu-Sato solution generally $P_3\not=m_3$ and $Q_3\not=q_3$, since $S_{10}=-H_{10}=im\hat{q}(b-\nu)$.

Looking at the lowest multipole moments, it is easy to see that the mass of the solution, the electric charge and the gravitational angular momentum are given by $m$, $\hat{q}$ and $ma$, respectively. The magnetic dipole moment is $m_d=-iP_1=\hat{q}(a-b+\nu)$. The mass-quadrupole moment is $P_2=-m\left(d-\delta+a^2-ab\right)$.

According to the results worked out in the papers \cite{Kordas95,Meinel95,Pachon06,Ernst06}, if the solution is reflection symmetric with respect to the equatorial plane, then $m_n$ is real for even $n$, and purely imaginary for odd $n$. The behavior of the constants $q_n$ is either the same as that of $m_n$, or just the opposite, $q_n$ is purely imaginary for even $n$, and real for odd $n$. Assuming that $\hat{q}$ is nonzero, the expression of the magnetic dipole moment shows that $\nu-b$ must be real for reflection symmetry. Using \eqref{eqdeld}, the quadrupole moment can be written into the alternative form $P_2=\frac{m}{2}\left(m^2+a^2-\hat{q}^2-b^2-2\kappa^2\right)$. From this follows that reflection symmetry is possible only if $b$ is either real or pure imaginary.

The multipole moments of the Kerr-Newman metric have been already presented in \cite{Sotiriou04} as $P_n=m_n=m(ia)^n$ and $Q_n=q_n=\hat{q}(ia)^n$. The quantities $M_{ij}$, $Q_{ij}$, $S_{ij}$, $H_{ij}$ defined in \eqref{eqmijsij}-\eqref{eqqijhij} are all identically zero in this case. From the expression \eqref{eqq1hatq} for $q_1$ follows that the 5-parameter solution can reduce to the Kerr-Newman metric only if $\nu=b$. Furthermore, comparing with the quadrupole moment $m_2$ follows that the solution becomes the Kerr-Newman metric only if $b^2=a^2+\hat{q}^2-m^2$. This shows that for the sub-extreme Kerr-Newman solution the parameters $b=\nu$ are purely imaginary.

Setting $b=0$ we obtain the charged magnetized generalization of the $\delta=2$ Tomimatsu-Sato solution published in \cite{Manko98,Manko99}. The original vacuum $\delta=2$ Tomimatsu-Sato solution can be obtained by setting $b=0$, $\hat q=0$ and $\mu=0$. The two parameters in the vacuum Tomimatsu-Sato solution, satisfying $p_{v}^{2}+q_{v}^{2}=1$, are related to the constants in the 5-parameter solution by $q_{v}=a/m$. Furthermore, $\kappa=mp_{v}/2$.

Instead of giving here the expressions for the first few multipole moments applying \eqref{eqp0m0}-\eqref{eqq6q6}, we continue by a direct calculation of the multipole moments, using a generalization of the method used in \cite{Hansen74,Backdahl05} to calculate the moments of the Kerr metric. That way we obtain a fast and efficient method by which one can obtain the multipole moments of the 5-parameter solution up to any desired order by a relatively simple algorithm. After the elimination of $\kappa$ using \eqref{eqkappasqrt}, the expansion coefficients $m_n$ and $q_n$ have identical form for the subextreme and the hyperextreme cases. This implies that the multipole moments are also the same for the two cases. For this reason it is sufficient to concentrate on the subextreme case in the following.

\subsection{Conformal mapping}

For the direct calculation of the multipole moments we also need the function $\gamma$ in the 3-metric \eqref{eqhabxy}. It would also be possible to calculate the multipole moments using the Weyl coordinates $\rho$ and $z$. However, this would make the expressions very complicated because of the  square roots appearing from the substitution \eqref{eqxyrpm} for $x$ and $y$. The square roots can be avoided by a different choice of asymptotic coordinates, and by a different conformal factor \cite{Hansen74,Backdahl05}. The first step is the introduction of asymptotic coordinates $\hat x^{\hat a}=(R,\theta,\phi)$ by
\begin{equation}
x=\frac{1}{\kappa R}+\frac{\kappa R}{4} \quad , \qquad
y=\cos\theta \ .
\end{equation}
In this case $R=0$ corresponds to conformal infinity $\Lambda$, and the nonzero components of the metric \eqref{eqhabxy} become
\begin{align}
h_{RR}=&\frac{e^{2\gamma}}{16R^4}
\left[16+\kappa^4R^4-8\kappa^2R^2\cos(2\theta)\right] \ , \\
h_{\theta\theta}=&\frac{e^{2\gamma}}{16R^2}
\left[16+\kappa^4R^4-8\kappa^2R^2\cos(2\theta)\right] \ , \\
h_{\phi\phi}=&\frac{1}{16R^2}\left(4-\kappa^2R^2\right)^2\sin^2\theta \ .
\end{align}
The clear advantage of this choice of $R$ coordinate is that $h_{\theta\theta}=R^2h_{RR}$. It follows that an appropriate choice of conformal factor is
\begin{equation}
\Omega=\frac{4R^2}{4-\kappa^2R^2} \ . \label{eqoomrr}
\end{equation}
The metric $\tilde h_{\hat a\hat b}=\Omega^2 h_{\hat a\hat b}$ can be extended smoothly to the point $\Lambda$. Introducing cylindrical coordinates $\tilde x^{\tilde a}=(\tilde\rho,\tilde z,\phi)$ by
\begin{equation}
R^2=\tilde\rho^2+\tilde z^2 \quad , \qquad
\cos\theta=\frac{\tilde z}{R} \ ,
\end{equation}
the metric takes the familiar form
\begin{equation}
\tilde h_{\tilde a\tilde b}=\left(
\begin{array}{ccc}
e^{2\tilde\gamma} & 0 & 0 \\
0 & e^{2\tilde\gamma} & 0 \\
0 & 0 & \tilde\rho^2
\end{array}
\right) \ ,
\end{equation}
where
\begin{equation}
e^{2\tilde\gamma}=e^{2\gamma}\left\{
1+\frac{16\kappa^2\tilde\rho^2}{\left[
	4-\kappa^2\left(\tilde\rho^2+\tilde z^2\right)
	\right]^2}\right\} \ . \label{eqe2game2gam}
\end{equation}
Comparing with \eqref{eqhabaxisyminf}, it might be surprising at first sight that a transformed version of the function $\gamma$ appears in the metric. The reason for this is that the coordinate $\tilde\rho$ and $\tilde z$ are clearly different now from the coordinates introduced earlier from the Weyl coordinates in \eqref{eqrhotilztil}. However, this representation of the conformal metric with the conformal factor $\Omega$ given in \eqref{eqoomrr} is just as appropriate for the calculation of the multipole moments as the Weyl coordinate approach, but it is without the emergence of the square roots.

In our case, the metric function $\gamma$ for the 5-parameter solution is given in \cite{Manko00b} as
\begin{equation}
e^{2\gamma}=\frac{E}{16\kappa^8\left(x^2-y^2\right)^4} \ , \label{eqe2gamee}
\end{equation}
where
\begin{align}
E=&\left\{4\left[\kappa^{2}\left(x^{2}-1\right)
+\delta\left(1-y^{2}\right)\right]^{2}
+c\left[c(d-\delta)-m^{2} b+\hat{q}^2 \nu\right]
\left(1-y^{2}\right)^{2}\right\}^{2} \nonumber \\
&-16 \kappa^{2}\left(x^{2}-1\right)\left(1-y^{2}\right)
\left\{c\left[\kappa^{2}\left(x^{2}-y^{2}\right)+2 \delta y^{2}\right]+\left(m^{2} b-\hat{q}^2 \nu\right) y^{2}\right\}^{2} \ . \label{eqeekappa}
\end{align}

\subsection{Calculation of the multipole moments}

We intend to calculate the necessary leading order functions and use the recursive formula \eqref{eqynind0}-\eqref{eqynind2} to calculate the multipole moments. We calculate the Ernst potentials $\xi$ and $q$ using their definition \eqref{eqxiqdef}, from the complex potentials $\mathcal{E}$ and $\Phi$ given by \eqref{eqephicmts}-\eqref{eqccxy}. Here we need to substitute
\begin{equation}
x=\frac{1}{\kappa R}+\frac{\kappa R}{4} \quad , \qquad
y=\frac{\tilde z}{R} \ ,
\end{equation}
where $R=\sqrt{\tilde\rho^2+\tilde z^2}$. Since we will take the leading order part, and since $R_L=0$, for the coordinate $x$ we can substitute the simpler expression $x=\frac{1}{\kappa R}$ without influencing the final result. According to \eqref{eqfldef}, the leading order part of a function can be obtained by substituting $\tilde z=\zeta$ and $\tilde\rho=-i\zeta$.

In order to start the recursion \eqref{eqynind0}-\eqref{eqynind2}, for the gravitational moments we need to calculate the leading order part of $\tilde\xi=\Omega^{-1/2}\xi$, and for the electromagnetic moments we need the leading order part of $\tilde q=\Omega^{-1/2}q$. Here we have to use the conformal factor given in \eqref{eqoomrr}. However, since we will take the leading order part, we can drop the $\kappa^2R^2$ term from the denominator of $\Omega$, and simply use $\Omega=R^2$ instead. The result for the leading order part of the conformally transformed complex potentials turns out to be
\begin{equation}
\tilde\xi_L=\frac{\alpha}{\beta} \quad , \qquad
\tilde q_L=\frac{\lambda}{\beta} \ ,
\end{equation}
where
\begin{align}
\alpha=&2 m \left[2+ (2 i-c \zeta)b \zeta-2 \delta  \zeta^2+i \zeta^3 \left(w-2b  \delta\right)\right]\ , \\
\beta=& w(c \zeta+4 i)\zeta^3 -4 i c \zeta+4 \left(1-\delta  \zeta^2\right)^2\ , \\
\lambda=&2 \hat{q} \left[2+ ( 2 i-c\zeta)\nu \zeta-2 \delta  \zeta^2+i \zeta^3 \left(w-2  \nu \delta\right)\right] \ .
\end{align}
and
\begin{equation}
w=vc + \hat{q}^2\nu - m^2b\quad , \qquad v=d-\delta \ .
\end{equation}
Since the leading order part of \eqref{eqe2game2gam} is simply $e^{2\tilde\gamma_L}=e^{2\gamma_L}\left(1-\kappa^2\zeta^2\right)$, using \eqref{eqe2gamee}-\eqref{eqeekappa} we obtain
\begin{equation}
e^{2\tilde\gamma_L}=
\frac{16 \left(c \zeta-w \zeta^3\right)^2+\left(c w \zeta^4+4 \left(1-\delta  \zeta^2\right)^2\right)^2}
{16 \left[1- (2 \delta +v)\zeta^2\right]^3} \ . \label{eqetilgaml}
\end{equation}
All these expressions determining $\tilde\xi_L$, $\tilde q_L$ and $\tilde\gamma_L$ are valid in both the subextreme and the hyperextreme cases.

We are now ready to define the functions $y_n$ according to \eqref{eqynind0}-\eqref{eqynind2}. For the gravitational moments we need to take $y_{0}=\tilde\xi_L$, and for the electromagnetic moments $y_{0}=\tilde q_L$. The function $y_{1}$ can be obtained by taking the $\zeta$ derivative of $y_{0}$, i.e.~$y_{1}=y_0{'}$. For the further $y_n$ functions we need to replace $\gamma$ by $\tilde\gamma$ in \eqref{eqynind2},
\begin{equation}
y_{n+1}=y_n{'}-2n\, y_n\tilde\gamma_L{'}
-\frac{1}{2}n(2n-1)\tilde R_L\,y_{n-1} \ . \label{eqynplus1}
\end{equation}
The $\zeta$ derivative of $\tilde\gamma_L$ can be obtained from \eqref{eqetilgaml}. Since it only depends on the structure of the 3-metric, equation \eqref{eqgammalpr} is valid now in terms of $\tilde\gamma$, giving $\tilde R_L=\frac{2}{\zeta}\tilde\gamma_L{'}$. The multipole moments can be obtained by taking the values of the $y_n$ functions at $\zeta=0$, according to \eqref{eqpnynfac}.

We present the results for the first few scalar moments by writing out the terms that must be added to the axis coefficients $m_n$ and $q_n$. These coefficients can be easily calculated for general $n$ according to \eqref{eqmkqkf}, and their values up to $n=5$ are listed in \eqref{eqm0cmts}-\eqref{eqq4cmts}. For the first six scalar multipole moments we obtain:
\begin{align}
P_0=&m_0 \ , \\
P_1=&m_1 \ , \\
P_2=&m_2 \ , \\
P_3=&m_3+\frac{ i m \hat{q}^2 }{5}(b-\nu)\ , \\
P_4=&m_4-\frac{m^3}{7}  (b^2+bc-v)-\frac{m \hat{q}^2}{35}
 \left(7 c b-12c\nu-8 b \nu+5 v+3 \nu^2\right) \ , \\
P_5=&m_5+\frac{im}{21}\Bigl\{
 \hat{q}^2\left(\hat{q}^2-m^2\right)(b-\nu)
 +m^2(b-6c)\left(b^2+bc-v\right) \\
 &+\hat{q}^2\left[c^2(11\nu-5b)-6cv+c\nu(8b-3\nu)+v(10\nu-9b)-b\nu^2
\right]\Bigr\} \ , \notag
\end{align}
and
\begin{align}
Q_0=&q_0 \ , \\
Q_1=&q_1 \ , \\
Q_2=&q_2 \ , \\
Q_3=&q_3-\frac{i\hat{q}m^2}{5}  (\nu-b) \ , \\
Q_4=&q_4+\dfrac{\hat{q}^3}{7} (\nu^2+\nu c-v)+\frac{\hat{q}m^2}{35}
 \left(7 c \nu-12cb-8 \nu b+5 v+3 b^2\right) \\
Q_5=&q_5+\frac{i\hat{q}}{21}\Bigl\{
 m^2\left(m^2-\hat{q}^2\right)(\nu-b)
 -\hat{q}^2(\nu-6c)\left(\nu^2+\nu c-v\right)\\
 &-m^2\left[c^2(11b-5\nu)-6cv+cb(8\nu-3b)+v(10b-9\nu)-\nu b^2
\right]\Bigr\} \notag \ .
\end{align}
We have already seen that if we make the replacements $m\leftrightarrow\hat{q}$, $b\leftrightarrow\nu$, keeping $c$ and $v$ fixed, then we obtain $q_n$ from $m_n$ and vice versa. However, the same formal symmetry is not valid between $P_n$ and $Q_n$, since some of the terms have opposite sign, even if they have the same form under this replacement. 

We note that the above forms of $P_n$ and $Q_n$ are not unique, since the variable $v$ is not independent from the constants $m$, $\hat{q}$, $b$, $\nu$ and $c$. The variable $v$ has been introduced in order to make the equations shorter by hiding the denominators. If we take the definition $v=d-\delta$, substitute $d$ and $\delta$ from \eqref{eqdeld}, move all terms to one side of the equation, and multiply with the denominator, we get
\begin{equation}
 v\left(m^2-c^2-\hat{q}^2\right)-\frac{1}{4}\left(m^2-c^2-\hat{q}^2\right)^2
 -m^2b^2+\hat{q}^2\nu^2=0 \ .  \label{eqvconstid}
\end{equation}
To all the obtained results for $P_n$ and $Q_n$ we can add this expression multiplied by an arbitrary polynomial, without changing their validity, and without introducing denominators. In order to obtain the above forms of $P_4$, $Q_4$, $P_5$ and $Q_5$ we have used \eqref{eqvconstid} to eliminate terms containing $c^2v$.

Higher order scalar moments can be relatively easily calculated by an algebraic manipulation software, calculating the functions $y_n$ according to \eqref{eqynplus1}. However, higher derivatives of $\tilde\xi_L$, $\tilde q_L$ and $\tilde\gamma_L$ become more and more complicated and their calculations can be very slow. If one is interested in the moments up to order $N$, then it is advisable to first take the power series expansion of the functions at $\zeta=0$ up to order $N$, and apply only then the recursion formula \eqref{eqynplus1}. That makes the calculations quite fast and memory efficient even up to high orders. We provide a Mathematica and an equivalent Maple file (named "five-par-sol") as supplementary material for the calculation of the moments of the 5-parameter solution.

\section*{Acknowledgement}

This study was financed in part by the Coordenação de Aperfeiçoamento de Pessoal de Nível Superior - Brasil (CAPES)-Finance Code 001. B. H. would like to thank FAPESP for financial support under grant 2019/01511-5 as well as the DFG Research Training Group 1620 Models of Gravity for financial support.

\bibliographystyle{hhieeetr}
\bibliography{electrovac}

\begin{thebibliography}{10}

\bibitem{Geroch70}
R.~Geroch, \href {http://dx.doi.org/10.1063/1.1665427} {``Multipole moments.
  {II.} {Curved} space,''} {\em Journal of Mathematical Physics}, vol.~11,
  no.~8, pp.~2580--2588, 1970.

\bibitem{Hansen74}
R.~O. Hansen, \href {http://dx.doi.org/10.1063/1.1666501} {``Multipole moments
  of stationary space‐times,''} {\em Journal of Mathematical Physics},
  vol.~15, no.~1, pp.~46--52, 1974.

\bibitem{Thorne80}
K.~S. Thorne, \href {http://dx.doi.org/10.1103/RevModPhys.52.299} {``Multipole
  expansions of gravitational radiation,''} {\em Rev. Mod. Phys.}, vol.~52,
  pp.~299--339, Apr 1980.

\bibitem{Gursel83}
Y.~G{\"u}rsel, \href {http://dx.doi.org/10.1007/BF01031881} {``Multipole
  moments for stationary systems: The equivalence of the {Geroch-Hansen}
  formulation and the {Thorne} formulation,''} {\em General Relativity and
  Gravitation}, vol.~15, no.~8, pp.~737--754, 1983.

\bibitem{Simon83}
W.~Simon and R.~Beig, \href {http://dx.doi.org/10.1063/1.525846} {``The
  multipole structure of stationary space‐times,''} {\em Journal of
  Mathematical Physics}, vol.~24, no.~5, pp.~1163--1171, 1983.

\bibitem{Quevedo90}
H.~Quevedo, \href {http://dx.doi.org/10.1002/prop.2190381002} {``Multipole
  moments in general relativity ---static and stationary vacuum
  solutions---,''} {\em Fortschritte der Physik/Progress of Physics}, vol.~38,
  no.~10, pp.~733--840, 1990.

\bibitem{Ryan95}
F.~D. Ryan, \href {http://dx.doi.org/10.1103/PhysRevD.52.5707} {``Gravitational
  waves from the inspiral of a compact object into a massive, axisymmetric body
  with arbitrary multipole moments,''} {\em Phys. Rev. D}, vol.~52,
  pp.~5707--5718, Nov 1995.

\bibitem{Ryan97}
F.~D. Ryan, \href {http://dx.doi.org/10.1103/PhysRevD.56.1845} {``Accuracy of
  estimating the multipole moments of a massive body from the gravitational
  waves of a binary inspiral,''} {\em Phys. Rev. D}, vol.~56, pp.~1845--1855,
  Aug 1997.

\bibitem{LiLovelace08}
C.~Li and G.~Lovelace, \href {http://dx.doi.org/10.1103/PhysRevD.77.064022}
  {``Generalization of {Ryan's} theorem: Probing tidal coupling with
  gravitational waves from nearly circular, nearly equatorial,
  extreme-mass-ratio inspirals,''} {\em Phys. Rev. D}, vol.~77, p.~064022, Mar
  2008.

\bibitem{Barack07}
L.~Barack and C.~Cutler, \href {http://dx.doi.org/10.1103/PhysRevD.75.042003}
  {``Using {LISA} extreme-mass-ratio inspiral sources to test {off-Kerr}
  deviations in the geometry of massive black holes,''} {\em Phys. Rev. D},
  vol.~75, p.~042003, Feb 2007.

\bibitem{Babak17}
S.~Babak {\em et~al.}, \href {http://dx.doi.org/10.1103/PhysRevD.95.103012}
  {``Science with the space-based interferometer {LISA}. {V}. {Extreme}
  mass-ratio inspirals,''} {\em Phys. Rev. D}, vol.~95, p.~103012, May 2017.

\bibitem{Chrusciel2012}
P.~T. Chru{\'s}ciel, J.~L. Costa, and M.~Heusler, \href
  {http://dx.doi.org/10.12942/lrr-2012-7} {``Stationary black holes:
  {Uniqueness} and beyond,''} {\em Living Reviews in Relativity}, vol.~15,
  no.~1, p.~7, 2012.

\bibitem{Cardoso16}
V.~Cardoso and L.~Gualtieri, \href
  {http://dx.doi.org/10.1088/0264-9381/33/17/174001} {``Testing the black hole
  `no-hair' hypothesis,''} {\em Classical and Quantum Gravity}, vol.~33,
  p.~174001, aug 2016.

\bibitem{Cardoso19}
V.~Cardoso and P.~Pani, \href {http://dx.doi.org/10.1007/s41114-019-0020-4}
  {``Testing the nature of dark compact objects: a status report,''} {\em
  Living Reviews in Relativity}, vol.~22, no.~1, p.~4, 2019.

\bibitem{Sopuerta09}
C.~F. Sopuerta and N.~Yunes, \href
  {http://dx.doi.org/10.1103/PhysRevD.80.064006} {``Extreme- and
  intermediate-mass ratio inspirals in dynamical chern-simons modified
  gravity,''} {\em Phys. Rev. D}, vol.~80, p.~064006, Sep 2009.

\bibitem{Pappas15a}
G.~Pappas and T.~P. Sotiriou, \href
  {http://dx.doi.org/10.1103/PhysRevD.91.044011} {``Multipole moments in
  scalar-tensor theory of gravity,''} {\em Phys. Rev. D}, vol.~91, p.~044011,
  Feb 2015.

\bibitem{Kleihaus14}
B.~Kleihaus, J.~Kunz, and S.~Mojica, \href
  {http://dx.doi.org/10.1103/PhysRevD.90.061501} {``Quadrupole moments of
  rapidly rotating compact objects in dilatonic {Einstein-Gauss-Bonnet}
  theory,''} {\em Phys. Rev. D}, vol.~90, p.~061501, Sep 2014.

\bibitem{Pappas15b}
G.~Pappas and T.~P. Sotiriou, \href {http://dx.doi.org/10.1093/mnras/stv1819}
  {``Geodesic properties in terms of multipole moments in scalar--tensor
  theories of gravity,''} {\em Monthly Notices of the Royal Astronomical
  Society}, vol.~453, pp.~2862--2876, 09 2015.

\bibitem{Collins04}
N.~A. Collins and S.~A. Hughes, \href
  {http://dx.doi.org/10.1103/PhysRevD.69.124022} {``Towards a formalism for
  mapping the spacetimes of massive compact objects: Bumpy black holes and
  their orbits,''} {\em Phys. Rev. D}, vol.~69, p.~124022, Jun 2004.

\bibitem{Glampedakis06}
K.~Glampedakis and S.~Babak, \href
  {http://dx.doi.org/10.1088/0264-9381/23/12/013} {``Mapping spacetimes with
  {LISA}: inspiral of a test body in a `quasi-{Kerr}' field,''} {\em Classical
  and Quantum Gravity}, vol.~23, pp.~4167--4188, may 2006.

\bibitem{Vigeland10}
S.~J. Vigeland, \href {http://dx.doi.org/10.1103/PhysRevD.82.104041}
  {``Multipole moments of bumpy black holes,''} {\em Phys. Rev. D}, vol.~82,
  p.~104041, Nov 2010.

\bibitem{Gair13}
J.~R. Gair, M.~Vallisneri, S.~L. Larson, and J.~G. Baker, \href
  {http://dx.doi.org/10.12942/lrr-2013-7} {``Testing general relativity with
  low-frequency, space-based gravitational-wave detectors,''} {\em Living
  Reviews in Relativity}, vol.~16, no.~1, p.~7, 2013.

\bibitem{Berti15}
E.~Berti {\em et~al.}, \href {http://dx.doi.org/10.1088/0264-9381/32/24/243001}
  {``Testing general relativity with present and future astrophysical
  observations,''} {\em Classical and Quantum Gravity}, vol.~32, p.~243001, dec
  2015.

\bibitem{Yagi16}
K.~Yagi and L.~C. Stein, \href
  {http://dx.doi.org/10.1088/0264-9381/33/5/054001} {``Black hole based tests
  of general relativity,''} {\em Classical and Quantum Gravity}, vol.~33,
  p.~054001, feb 2016.

\bibitem{Will08}
C.~M. Will, \href {http://dx.doi.org/10.1086/528847} {``Testing the general
  relativistic "no-hair" theorems using the galactic center black hole
  {Sagittarius} {$A^{*}$},''} {\em The Astrophysical Journal}, vol.~674,
  pp.~L25--L28, jan 2008.

\bibitem{Broderick14}
A.~E. Broderick, T.~Johannsen, A.~Loeb, and D.~Psaltis, \href
  {http://dx.doi.org/10.1088/0004-637x/784/1/7} {``Testing the no-hair theorem
  with event horizon telescope observations of {Sagittarius} {$A^{*}$},''} {\em
  The Astrophysical Journal}, vol.~784, p.~7, feb 2014.

\bibitem{Suvorov16}
A.~G. Suvorov and A.~Melatos, \href
  {http://dx.doi.org/10.1103/PhysRevD.93.024004} {``Testing modified gravity
  and no-hair relations for the {Kerr-Newman} metric through quasiperiodic
  oscillations of galactic microquasars,''} {\em Phys. Rev. D}, vol.~93,
  p.~024004, Jan 2016.

\bibitem{Psaltis16}
D.~Psaltis, N.~Wex, and M.~Kramer, \href
  {http://dx.doi.org/10.3847/0004-637x/818/2/121} {``A quantitative test of the
  no-hair theorem with {Sgr} {$A^{*}$} using stars, pulsars, and the event
  horizon telescope,''} {\em The Astrophysical Journal}, vol.~818, p.~121, feb
  2016.

\bibitem{Paschalidis17}
V.~Paschalidis and N.~Stergioulas, \href
  {http://dx.doi.org/10.1007/s41114-017-0008-x} {``Rotating stars in
  relativity,''} {\em Living Reviews in Relativity}, vol.~20, no.~1, p.~7,
  2017.

\bibitem{Maselli20}
A.~Maselli, G.~Pappas, P.~Pani, L.~Gualtieri, S.~Motta, V.~Ferrari, and
  L.~Stella, \href {http://dx.doi.org/10.3847/1538-4357/ab9ff4} {``A new method
  to constrain neutron star structure from quasi-periodic oscillations,''} {\em
  The Astrophysical Journal}, vol.~899, p.~139, aug 2020.

\bibitem{Shibata98}
M.~Shibata and M.~Sasaki, \href {http://dx.doi.org/10.1103/PhysRevD.58.104011}
  {``Innermost stable circular orbits around relativistic rotating stars,''}
  {\em Phys. Rev. D}, vol.~58, p.~104011, Oct 1998.

\bibitem{SanabriaGomez10}
J.~D. Sanabria-G{\'o}mez, J.~L. Hern{\'a}ndez-Pastora, and F.~L. Dubeibe, \href
  {http://dx.doi.org/10.1103/PhysRevD.82.124014} {``Innermost stable circular
  orbits around magnetized rotating massive stars,''} {\em Phys. Rev. D},
  vol.~82, p.~124014, Dec 2010.

\bibitem{Berti04}
E.~Berti and N.~Stergioulas, \href
  {http://dx.doi.org/10.1111/j.1365-2966.2004.07740.x} {``Approximate matching
  of analytic and numerical solutions for rapidly rotating neutron stars,''}
  {\em Monthly Notices of the Royal Astronomical Society}, vol.~350,
  pp.~1416--1430, 06 2004.

\bibitem{Pappas14}
G.~Pappas and T.~A. Apostolatos, \href
  {http://dx.doi.org/10.1103/PhysRevLett.112.121101} {``Effectively universal
  behavior of rotating neutron stars in general relativity makes them even
  simpler than their newtonian counterparts,''} {\em Phys. Rev. Lett.},
  vol.~112, p.~121101, Mar 2014.

\bibitem{Yagi14}
K.~Yagi, K.~Kyutoku, G.~Pappas, N.~Yunes, and T.~A. Apostolatos, \href
  {http://dx.doi.org/10.1103/PhysRevD.89.124013} {``Effective no-hair relations
  for neutron stars and quark stars: Relativistic results,''} {\em Phys. Rev.
  D}, vol.~89, p.~124013, Jun 2014.

\bibitem{Yagi17}
K.~Yagi and N.~Yunes, \href {http://dx.doi.org/10.1016/j.physrep.2017.03.002}
  {``Approximate universal relations for neutron stars and quark stars,''} {\em
  Physics Reports}, vol.~681, pp.~1--72, 2017.
\newblock Approximate Universal Relations for Neutron Stars and Quark Stars.

\bibitem{Manko95}
V.~S. Manko, J.~Mart{\'\i}n, and E.~Ruiz, \href
  {http://dx.doi.org/10.1063/1.531012} {``Six‐parameter solution of the
  {Einstein--Maxwell} equations possessing equatorial symmetry,''} {\em Journal
  of Mathematical Physics}, vol.~36, no.~6, pp.~3063--3073, 1995.

\bibitem{Manko00}
V.~S. Manko, E.~W. Mielke, and J.~D. Sanabria-G{\'o}mez, \href
  {http://dx.doi.org/10.1103/PhysRevD.61.081501} {``Exact solution for the
  exterior field of a rotating neutron star,''} {\em Phys. Rev. D}, vol.~61,
  p.~081501, Mar 2000.

\bibitem{Manko00b}
V.~S. Manko, J.~D. Sanabria-G{\'o}mez, and O.~V. Manko, \href
  {http://dx.doi.org/10.1103/PhysRevD.62.044048} {``Nine-parameter electrovac
  metric involving rational functions,''} {\em Phys. Rev. D}, vol.~62,
  p.~044048, Jul 2000.

\bibitem{Teichmuller11}
C.~Teichm{\"u}ller, M.~B. Fr{\"o}b, and F.~Maucher, \href
  {http://dx.doi.org/10.1088/0264-9381/28/15/155015} {``Analytical
  approximation of the exterior gravitational field of rotating neutron
  stars,''} {\em Classical and Quantum Gravity}, vol.~28, p.~155015, jul 2011.

\bibitem{Pachon12}
L.~A. Pach{\'o}n, J.~A. Rueda, and J.~D. Sanabria-G{\'o}mez, \href
  {http://dx.doi.org/10.1103/PhysRevD.73.104038} {``Realistic exact solution
  for the exterior field of a rotating neutron star,''} {\em Phys. Rev. D},
  vol.~73, p.~104038, May 2006.

\bibitem{Pappas13}
G.~Pappas and T.~A. Apostolatos, \href {http://dx.doi.org/10.1093/mnras/sts556}
  {``An all-purpose metric for the exterior of any kind of rotating neutron
  star,''} {\em Monthly Notices of the Royal Astronomical Society}, vol.~429,
  pp.~3007--3024, 01 2013.

\bibitem{Manko16}
V.~S. Manko and E.~Ruiz, \href {http://dx.doi.org/10.1103/PhysRevD.93.104051}
  {``Exterior field of slowly and rapidly rotating neutron stars:
  Rehabilitating spacetime metrics involving hyperextreme objects,''} {\em
  Phys. Rev. D}, vol.~93, p.~104051, May 2016.

\bibitem{Xanthopoulos79}
B.~C. Xanthopoulos, \href {http://dx.doi.org/10.1088/0305-4470/12/7/018}
  {``Multipole moments in general relativity,''} {\em Journal of Physics A:
  Mathematical and General}, vol.~12, pp.~1025--1028, jul 1979.

\bibitem{Beig80}
R.~Beig and W.~Simon, \href {http://dx.doi.org/10.1007/BF01941970} {``Proof of
  a multipole conjecture due to {Geroch},''} {\em Communications in
  Mathematical Physics}, vol.~78, no.~1, pp.~75--82, 1980.

\bibitem{Kundu81}
P.~Kundu, \href {http://dx.doi.org/10.1063/1.525148} {``On the analyticity of
  stationary gravitational fields at spatial infinity,''} {\em Journal of
  Mathematical Physics}, vol.~22, no.~9, pp.~2006--2011, 1981.

\bibitem{Beig81}
R.~Beig, W.~Simon, and R.~Penrose, \href
  {http://dx.doi.org/10.1098/rspa.1981.0095} {``On the multipole expansion for
  stationary space-times,''} {\em Proceedings of the Royal Society of London.
  A. Mathematical and Physical Sciences}, vol.~376, no.~1765, pp.~333--341,
  1981.

\bibitem{Hoenselaers86}
C.~Hoenselaers, ``On multipole moments in general relativity,'' in {\em
  Gravitational Collapse and Relativity: Proceedings of Yamada Conference XIV}
  (H.~Sato and T.~Nakamura, eds.), pp.~176--184, World Scientific, 1986.

\bibitem{Fodor89}
G.~Fodor, C.~Hoenselaers, and Z.~Perj{\'e}s, \href
  {http://dx.doi.org/10.1063/1.528551} {``Multipole moments of axisymmetric
  systems in relativity,''} {\em Journal of Mathematical Physics}, vol.~30,
  no.~10, pp.~2252--2257, 1989.

\bibitem{Kleinwachter95}
A.~Kleinw{\"a}chter, R.~Meinel, and G.~Neugebauer, \href
  {http://dx.doi.org/10.1016/0375-9601(95)00165-Y} {``The multipole moments of
  the rigidly rotating disk of dust in general relativity,''} {\em Physics
  Letters A}, vol.~200, no.~2, pp.~82--86, 1995.

\bibitem{Kordas95}
P.~Kordas, \href {http://dx.doi.org/10.1088/0264-9381/12/8/018}
  {``Reflection-symmetric, asymptotically flat solutions of the vacuum
  axistationary {Einstein} equations,''} {\em Classical and Quantum Gravity},
  vol.~12, pp.~2037--2044, aug 1995.

\bibitem{Meinel95}
R.~Meinel and G.~Neugebauer, \href
  {http://dx.doi.org/10.1088/0264-9381/12/8/019} {``Asymptotically flat
  solutions to the {Ernst} equation with reflection symmetry,''} {\em Classical
  and Quantum Gravity}, vol.~12, pp.~2045--2050, aug 1995.

\bibitem{Herberthson04}
M.~Herberthson, \href {http://dx.doi.org/10.1088/0264-9381/21/22/007} {``The
  gravitational dipole and explicit multipole moments of static axisymmetric
  spacetimes,''} {\em Classical and Quantum Gravity}, vol.~21, pp.~5121--5138,
  oct 2004.

\bibitem{Backdahl05a}
T.~B{\"a}ckdahl and M.~Herberthson, \href
  {http://dx.doi.org/10.1088/0264-9381/22/9/009} {``Static axisymmetric
  spacetimes with prescribed multipole moments,''} {\em Classical and Quantum
  Gravity}, vol.~22, pp.~1607--1621, apr 2005.

\bibitem{Backdahl05}
T.~B{\"a}ckdahl and M.~Herberthson, \href
  {http://dx.doi.org/10.1088/0264-9381/22/17/017} {``Explicit multipole moments
  of stationary axisymmetric spacetimes,''} {\em Classical and Quantum
  Gravity}, vol.~22, pp.~3585--3594, aug 2005.

\bibitem{Backdahl07}
T.~B{\"a}ckdahl, \href {http://dx.doi.org/10.1088/0264-9381/24/9/004}
  {``Axisymmetric stationary solutions with arbitrary multipole moments,''}
  {\em Classical and Quantum Gravity}, vol.~24, pp.~2205--2215, apr 2007.

\bibitem{Backdahl06}
T.~B{\"a}ckdahl and M.~Herberthson, \href
  {http://dx.doi.org/10.1088/0264-9381/23/20/019} {``Calculation of, and bounds
  for, the multipole moments of stationary spacetimes,''} {\em Classical and
  Quantum Gravity}, vol.~23, pp.~5997--6006, sep 2006.

\bibitem{Herberthson09}
M.~Herberthson, \href {http://dx.doi.org/10.1088/0264-9381/26/21/215009}
  {``Static spacetimes with prescribed multipole moments: a proof of a
  conjecture by {Geroch},''} {\em Classical and Quantum Gravity}, vol.~26,
  p.~215009, oct 2009.

\bibitem{Simon84}
W.~Simon, \href {http://dx.doi.org/10.1063/1.526271} {``The multipole expansion
  of stationary {Einstein--Maxwell} fields,''} {\em Journal of Mathematical
  Physics}, vol.~25, no.~4, pp.~1035--1038, 1984.

\bibitem{Pachon06}
L.~A. Pach{\'o}n and J.~D. Sanabria-G{\'o}mez, \href
  {http://dx.doi.org/10.1088/0264-9381/23/9/n01} {``Note on reflection symmetry
  in stationary axisymmetric electrovacuum spacetimes,''} {\em Classical and
  Quantum Gravity}, vol.~23, pp.~3251--3254, apr 2006.

\bibitem{Ernst06}
F.~J. Ernst, V.~S. Manko, and E.~Ruiz, \href
  {http://dx.doi.org/10.1088/0264-9381/23/15/013} {``Equatorial
  symmetry/antisymmetry of stationary axisymmetric electrovac spacetimes,''}
  {\em Classical and Quantum Gravity}, vol.~23, pp.~4945--4952, jul 2006.

\bibitem{Hoenselaers90}
C.~Hoenselaers and Z.~Perj{\'e}s, \href
  {http://dx.doi.org/10.1088/0264-9381/7/10/012} {``Multipole moments of
  axisymmetric electrovacuum spacetimes,''} {\em Classical and Quantum
  Gravity}, vol.~7, pp.~1819--1825, oct 1990.

\bibitem{Sotiriou04}
T.~P. Sotiriou and T.~A. Apostolatos, \href
  {http://dx.doi.org/10.1088/0264-9381/21/24/003} {``Corrections and comments
  on the multipole moments of axisymmetric electrovacuum spacetimes,''} {\em
  Classical and Quantum Gravity}, vol.~21, pp.~5727--5733, nov 2004.

\bibitem{Perjes03}
Z.~Perj{\'e}s, \href {http://dx.doi.org/10.1142/9789812704030_0006}
  {``Gravitational multipole moments,''} in {\em The Tenth Marcel Grossmann
  Meeting (2003)} (M.~Novello, S.~P. Bergliaffa, and R.~Ruffini, eds.),
  pp.~59--69, World Scientific, 2006.

\bibitem{Berti05}
E.~Berti, F.~White, A.~Maniopoulou, and M.~Bruni, \href
  {http://dx.doi.org/10.1111/j.1365-2966.2005.08812.x} {``Rotating neutron
  stars: an invariant comparison of approximate and numerical space--time
  models,''} {\em Monthly Notices of the Royal Astronomical Society}, vol.~358,
  pp.~923--938, 04 2005.

\bibitem{Ernst68a}
F.~J. Ernst, \href {http://dx.doi.org/10.1103/PhysRev.167.1175} {``New
  formulation of the axially symmetric gravitational field problem,''} {\em
  Phys. Rev.}, vol.~167, pp.~1175--1178, Mar 1968.

\bibitem{Ernst68b}
F.~J. Ernst, \href {http://dx.doi.org/10.1103/PhysRev.168.1415} {``New
  formulation of the axially symmetric gravitational field problem. {II},''}
  {\em Phys. Rev.}, vol.~168, pp.~1415--1417, Apr 1968.

\bibitem{Harrison68}
B.~K. Harrison, \href {http://dx.doi.org/10.1063/1.1664508} {``New solutions of
  the {Einstein‐Maxwell} equations from old,''} {\em Journal of Mathematical
  Physics}, vol.~9, no.~11, pp.~1744--1752, 1968.

\bibitem{Ernst71}
F.~J. Ernst, \href {http://dx.doi.org/10.1063/1.1665549}
  {``Exterior‐algebraic derivation of {Einstein} field equations employing a
  generalized basis,''} {\em Journal of Mathematical Physics}, vol.~12, no.~11,
  pp.~2395--2397, 1971.

\bibitem{Israel72}
W.~Israel and G.~A. Wilson, \href {http://dx.doi.org/10.1063/1.1666066} {``A
  class of stationary electromagnetic vacuum fields,''} {\em Journal of
  Mathematical Physics}, vol.~13, no.~6, pp.~865--867, 1972.

\bibitem{Kinnersley73}
W.~Kinnersley, \href {http://dx.doi.org/10.1063/1.1666373} {``Generation of
  stationary {Einstein‐Maxwell} fields,''} {\em Journal of Mathematical
  Physics}, vol.~14, no.~5, pp.~651--653, 1973.

\bibitem{Geroch71}
R.~Geroch, \href {http://dx.doi.org/10.1063/1.1665681} {``A method for
  generating solutions of {Einstein's} equations,''} {\em Journal of
  Mathematical Physics}, vol.~12, no.~6, pp.~918--924, 1971.

\bibitem{Stephani03}
H.~Stephani, D.~Kramer, M.~MacCallum, C.~Hoenselaers, and E.~Herlt, \href
  {http://dx.doi.org/10.1017/CBO9780511535185} {{\em Exact Solutions of
  Einstein's Field Equations}}.
\newblock Cambridge Monographs on Mathematical Physics, Cambridge University
  Press, 2~ed., 2003.

\bibitem{Penrose65}
R.~Penrose, \href {http://dx.doi.org/10.1098/rspa.1965.0058} {``Zero rest-mass
  fields including gravitation: asymptotic behaviour,''} {\em Proceedings of
  the Royal Society of London. Series A. Mathematical and Physical Sciences},
  vol.~284, no.~1397, pp.~159--203, 1965.

\bibitem{Trautman65}
A.~Trautman, ``Introduction to gravitational radiation theory,'' in {\em
  Lectures on General Relativity} (H.~Bondi, A.~Trautman, and F.~A.~E. Pirani,
  eds.), Brandeis Summer Institute in Theoretical Physics, pp.~249--373,
  Prentice-Hall, 1965.

\bibitem{Sotiriou05}
T.~P. Sotiriou and T.~A. Apostolatos, \href
  {http://dx.doi.org/10.1103/PhysRevD.71.044005} {``Tracing the geometry around
  a massive, axisymmetric body to measure, through gravitational waves, its
  mass moments and electromagnetic moments,''} {\em Phys. Rev. D}, vol.~71,
  p.~044005, Feb 2005.

\bibitem{Beig81a}
R.~Beig, \href {https://inis.iaea.org/search/search.aspx?orig_q=RN:13654481}
  {``The multipole expansion in general relativity,''} {\em Acta Physica
  Austriaca}, vol.~53, no.~4, pp.~249--270, 1981.

\bibitem{Fodor91}
G.~Fodor, \href {http://www.rmki.kfki.hu/~gfodor/Fodor92.pdf} {``Gravitational
  multipole moments,''} in {\em Relativity Today} (Z.~Perj{\'e}s, ed.),
  pp.~121--131, Nova Science Publishers, 1991.
\newblock Proceedings of the Third Hungarian Workshop, 1989 Tihany, ISBN
  156072028X,\ \url{http://www.rmki.kfki.hu/~gfodor/Fodor92.pdf}.

\bibitem{Manko95b}
E.~Ruiz, V.~S. Manko, and J.~Mart{\'\i}n, \href
  {http://dx.doi.org/10.1103/PhysRevD.51.4192} {``Extended {N-soliton} solution
  of the {Einstein-Maxwell} equations,''} {\em Phys. Rev. D}, vol.~51,
  pp.~4192--4197, Apr 1995.

\bibitem{tomimatsu72}
A.~Tomimatsu and H.~Sato, \href {http://dx.doi.org/10.1103/PhysRevLett.29.1344}
  {``New exact solution for the gravitational field of a spinning mass,''} {\em
  Phys. Rev. Lett.}, vol.~29, pp.~1344--1345, Nov 1972.

\bibitem{tomimatsu73}
A.~Tomimatsu and H.~Sato, \href {http://dx.doi.org/10.1143/PTP.50.95} {``New
  series of exact solutions for gravitational fields of spinning masses,''}
  {\em Progress of Theoretical Physics}, vol.~50, pp.~95--110, 07 1973.

\bibitem{Yamazaki77a}
M.~Yamazaki, \href {http://dx.doi.org/10.1143/PTP.57.1951} {``On the {Kerr} and
  the {Tomimatsu-Sato} spinning mass solutions,''} {\em Progress of Theoretical
  Physics}, vol.~57, pp.~1951--1957, 06 1977.

\bibitem{Yamazaki77b}
M.~Yamazaki, \href {http://dx.doi.org/10.1063/1.523213} {``On the
  {Kerr--Tomimatsu--Sato} family of spinning mass solutions,''} {\em Journal of
  Mathematical Physics}, vol.~18, no.~12, pp.~2502--2508, 1977.

\bibitem{Ernst73}
F.~J. Ernst, \href {http://dx.doi.org/10.1103/PhysRevD.7.2520} {``Charged
  version of {Tomimatsu-Sato} spinning-mass field,''} {\em Phys. Rev. D},
  vol.~7, pp.~2520--2521, Apr 1973.

\bibitem{Yamazaki78}
M.~Yamazaki, \href {http://dx.doi.org/10.1063/1.523835} {``On the charged
  {Kerr--Tomimatsu--Sato} family of solutions,''} {\em Journal of Mathematical
  Physics}, vol.~19, no.~6, pp.~1376--1378, 1978.

\bibitem{Manko98}
O.~V. Manko, V.~S. Manko, and J.~D. Sanabria-G{\'o}mez, \href
  {http://dx.doi.org/10.1143/PTP.100.671} {``Charged, magnetized
  {Tomimatsu-Sato} {$\delta=2$} solution,''} {\em Progress of Theoretical
  Physics}, vol.~100, pp.~671--673, 09 1998.

\bibitem{Manko99}
O.~V. Manko, V.~S. Manko, and J.~D. Sanabria-G{\'o}mez, \href
  {http://dx.doi.org/10.1023/A:1026782404418} {``Remarks on the charged,
  magnetized {Tomimatsu-Sato} {$\delta=2$} solution,''} {\em General Relativity
  and Gravitation}, vol.~31, no.~10, pp.~1539--1548, 1999.

\bibitem{Kinnersley74}
W.~Kinnersley and E.~F. Kelley, \href {http://dx.doi.org/10.1063/1.1666592}
  {``Limits of the {Tomimatsu‐Sato} gravitational field,''} {\em Journal of
  Mathematical Physics}, vol.~15, no.~12, pp.~2121--2126, 1974.

\bibitem{Tanabe76}
Y.~Tanabe, \href {http://dx.doi.org/10.1143/PTP.55.106} {``Multipole moments in
  general relativity,''} {\em Progress of Theoretical Physics}, vol.~55,
  pp.~106--114, 01 1976.

\bibitem{Kramer80}
D.~Kramer and G.~Neugebauer, \href
  {http://dx.doi.org/10.1016/0375-9601(80)90556-3} {``The superposition of two
  {Kerr} solutions,''} {\em Physics Letters A}, vol.~75, no.~4, pp.~259--261,
  1980.

\bibitem{Manko97}
V.~S. Manko and C.~Moreno, \href {http://dx.doi.org/10.1142/S0217732397000637}
  {``Extension of the parameter space in the {Tomimatsu--Sato} solutions,''}
  {\em Modern Physics Letters A}, vol.~12, no.~09, pp.~613--617, 1997.

\bibitem{Manko13}
V.~Manko and E.~Ruiz, \href {http://dx.doi.org/10.1093/ptep/ptt079}
  {``Singularities in the {Kerr--Newman} and charged {$\delta=2$}
  {Tomimatsu--Sato} spacetimes endowed with negative mass,''} {\em Progress of
  Theoretical and Experimental Physics}, vol.~2013, 10 2013.
\newblock 103E01.

\bibitem{Rosquist08}
K.~Rosquist, \href {http://dx.doi.org/10.1142/9789812834300_0391} {``Some
  physical consequences of the multipole structure of the {Kerr and
  Kerr-Newman} solutions,''} in {\em The Eleventh Marcel Grossmann Meeting}
  (H.~Kleinert, R.~T. Jantzen, and R.~Ruffini, eds.), pp.~2294--2298, World
  Scientific, 2008.

\bibitem{Dietz85}
W.~Dietz and C.~Hoenselaers, \href
  {http://dx.doi.org/10.1016/0003-4916(85)90301-X} {``Two mass solutions of
  {Einstein's} vacuum equations: The double {Kerr} solution,''} {\em Annals of
  Physics}, vol.~165, no.~2, pp.~319--383, 1985.

\bibitem{Kleinwachter11}
A.~Kleinw{\"a}chter, H.~Labranche, and R.~Meinel, \href
  {http://dx.doi.org/10.1007/s10714-010-1135-9} {``On the black hole limit of
  rotating discs and rings,''} {\em General Relativity and Gravitation},
  vol.~43, no.~5, pp.~1469--1486, 2011.

\bibitem{Breithaupt15}
M.~Breithaupt, Y.-C. Liu, R.~Meinel, and S.~Palenta, \href
  {http://dx.doi.org/10.1088/0264-9381/32/13/135022} {``On the black hole limit
  of rotating discs of charged dust,''} {\em Classical and Quantum Gravity},
  vol.~32, p.~135022, jun 2015.

\bibitem{Felice82}
F.~de~Felice and Y.~Yunqiang, \href
  {http://dx.doi.org/10.1088/0305-4470/15/10/036} {``Stellar rotation and
  gravitational collapse: the a/m issue,''} {\em Journal of Physics A:
  Mathematical and General}, vol.~15, pp.~3341--3349, oct 1982.

\bibitem{Bradley91}
M.~{Bradley}, A.~{Curir}, and F.~{de Felice}, \href
  {http://dx.doi.org/10.1086/170629} {``Massive black holes in the nuclei of
  spiral galaxies: M31 and the problem of the initial conditions,''} {\em
  Astrophysical Journal}, vol.~381, p.~72, 1991.

\end{thebibliography}

\end{document}